\documentclass[aps,pra,twocolumn,superscriptaddress,longbibliography,reprint]{revtex4-1}
\setcitestyle{square}

\usepackage{graphicx}
\usepackage{enumerate}
\usepackage[colorlinks=true, linkcolor=blue, citecolor=blue]{hyperref}
\usepackage{textcomp}
\usepackage{amsmath}
\usepackage{amssymb}
\usepackage[english]{babel}
\usepackage[capitalise]{cleveref}

\def\ud{\mathrm{d}}

\newcommand{\nep}{\textrm{e}}

\newcommand{\chain}{\mathrm{c} }
\newcommand{\Hc}{\mathrm{H.c.} }

\newcommand{\AIM}{\scriptscriptstyle \mathrm{SIAM}} 
\newcommand{\SIAM}{\scriptscriptstyle \mathrm{SIAM}}

\newcommand{\Tr}{{\rm Tr}}

\newcommand{\opa}[1]{{\hat{a}^{\phantom \dagger}}_{#1}}
\newcommand{\opadag}[1]{{\hat{a}^{\dagger}}_{#1}}
\newcommand{\opc}[1]{{\hat{c}^{\phantom \dagger}}_{#1}}
\newcommand{\opcdag}[1]{{\hat{c}^{\dagger}}_{#1}}
\newcommand{\opd}[1]{{\hat{d}^{\phantom \dagger}}_{#1}}
\newcommand{\opddag}[1]{{\hat{d}^{\dagger}}_{#1}}
\newcommand{\opfdag}[1]{{\hat{f}^{\dagger}}_{#1}}
\newcommand{\opf}[1]{{\hat{f}^{\phantom \dagger}}_{#1}}
\newcommand{\opn}[1]{{\hat{n}^{\phantom \dagger}}_{#1}}

\newcommand{\Ham}{\widehat{H}}
\newcommand{\Hloc}{\widehat{H}_{\mathrm{loc}}}
\newcommand{\Hhyb}{\widehat{H}_{\mathrm{hyb}}}
\newcommand{\Hcond}{\widehat{H}_{\mathrm{cond}}}

\newcommand{\fullemptyKet}{|\emptyset_1, {\mathrm F}_2\rangle}

\begin{document}

\title{Quenching the Anderson impurity model at finite temperature:\\ Entanglement and bath dynamics using matrix product states}
\author{Lucas Kohn}
\affiliation{SISSA, Via Bonomea 265, I-34136 Trieste, Italy}
\author{Giuseppe E. Santoro}
\affiliation{SISSA, Via Bonomea 265, I-34136 Trieste, Italy}
\affiliation{International Centre for Theoretical Physics (ICTP), P.O.Box 586, I-34014 Trieste, Italy}
\affiliation{CNR-IOM, Consiglio Nazionale delle Ricerche - Istituto Officina dei Materiali, c/o SISSA Via Bonomea 265, 34136 Trieste, Italy}

\begin{abstract}
We study the dynamics of the quenched Anderson model at finite temperature using matrix product states. Exploiting a chain mapping for the electron bath, we investigate the entanglement structure in the MPS for various orderings of the two chains, which emerge from the thermofield transformation employed to deal with nonzero temperature. We show that merging both chains can significantly lower the entanglement at finite temperatures as compared to an intuitive nearest-neighbor implementation of the Hamiltonian.
Analyzing the population of the free bath modes --possible when simulating the full dynamics of impurity plus bath -- we find clear signatures of the Kondo effect in the quench dynamics.
\end{abstract}

\maketitle

\section{Introduction}

The Anderson model is one of the most prominent models in condensed matter physics. Introduced to study the effect of a magnetic impurity in a metal \cite{Anderson_PR61}, and the emerging Kondo effect~\cite{Kondo_PTP64,Hewson_kondo:book}, it finds applications in various fields. 
In the framework of dynamical mean-field theory (DMFT), the single-impurity Anderson model (SIAM) serves as a building block to study strongly correlated materials~\cite{Metzner_PRL89,Georges_RMP96}. 
Coupled to an additional bath, the SIAM provides a simple playground to study quantum transport through the impurity, induced by a temperature gradient or an electric voltage between the leads \cite{PRL_Rams_2020,PRL_Weichselbaum_2018}, as realizable with quantum dots \cite{RMP_Reimann_2002,PRB_Braun_2004}.

The most popular techniques used so far to study the Anderson model are exact diagonalization (ED), Quantum Monte Carlo~\cite{RMP_Gull_2011, PRB_Rubtsov_2005, PRL_Werner_2006}, Wilson's numerical renormalization group (NRG)~\cite{Wilson_RMP75, RMP_Bulla_2008, PRL_Stadler_2015, PRL_Bulla_1999, PRB_Zitko_2009, PRL_Zitko_2013} and tensor-network-based methods \cite{PRL_Hallberg_2004,Wolf_PRB14,PRB_Wolf_2014_Cheb,PRX_Wolf_2015,PRB_Verstraete_2015,PRX_Bauernfeind_2017,PRL_Weichselbaum_2018,PRB_Linden_2020}, all of them with their own advantages and disadvantages.
ED is numerically exact and has equal resolution on all energy scales, but is usually limited in the number of conduction modes that can be treated. Monte-Carlo-based methods and NRG are very successful in calculating equilibrium properties, such as the impurity Green's function for DMFT applications. However, simulating real-time dynamics is more challenging. 
Matrix product states (MPS), or tensor-network methods in general, are well suited to deal with one dimensional systems~\cite{Schollwoeck_RMP05,AoP_Schollwoeck_2011}:
ground states of 1D models with short range interactions are known to follow an area law for the entanglement entropy~\cite{RMP_Plenio_2010}, making tensor networks a very efficient tool for equilibrium simulations. However, when simulating the dynamics, e.g. after a sudden quench, the entanglement typically grows in time, often even linearly. This results in an exponential increase of the required numerical resources. 


For real-time simulations employing MPS, it is crucial to reduce the entanglement as much as possible. For the Anderson model, a very natural idea to represent the conduction modes would be to apply a Wilson's chain mapping~\cite{Wilson_RMP75}: free electrons are represented by a tight-binding chain, which is well suited for NRG calculations and was believed to be the best strategy for MPS simulations as well, due to the interactions being short range. 
However, it has been shown that simulations in the so called ``star-geometry'', avoiding the Wilson's chain mapping, show significantly less entanglement~\cite{Wolf_PRB14}.

The problem with the standard chain mapping is easy to understand by considering the conduction electrons in absence of impurity.
The $T=0$ ground state is a simple product state in the star geometry, with modes below the Fermi energy $\epsilon_f$ occupied, and modes above $\epsilon_f$ empty. The standard chain mapping, on the contrary, is mixing-up all modes, leading to partially occupied chain sites, with nonzero entanglement even in the decoupled ground state. 
In a recent paper we have shown that the entanglement in the chain structure is significantly reduced by separating filled and empty modes~\cite{ArXiv_Kohn_2020}, mapping them into two independent chains. This approach shows low entanglement in the MPS, and also neatly generalizes to finite temperatures, by using the thermofield transformation: in contrast to the original matrix product density-operator-based approach~\cite{PRL_Cirac_2004,PRL_Vidal_2004}, it does not require imaginary-time evolution to deal with nonzero temperatures. 

The present paper discusses several different orderings of the sites of the improved chain mapping presented in \cite{ArXiv_Kohn_2020}, which one can still arbitrarily choose in setting-up an MPS encoding of the resulting nearest-neighbor Hamiltonian. 
We will show that the entanglement growth can be significantly reduced at finite temperature, allowing for much longer simulations, by
appropriately alternating filled and empty chain sites in the MPS. 
%
%



The paper is organized as follows. In \cref{sec:model} we introduce the model, and summarize our approach based on an improved chain mapping~\cite{ArXiv_Kohn_2020}, including the thermofield transformation to work at finite temperature.
We further discuss the different possible ordering of sites in the MPS that we have considered. 
\cref{sec:results} illustrates the results we have obtained, concerning in particular the SIAM in the Kondo regime.
To keep the discussion simple, we consider a quantum-quench scenario, where impurity and thermal conduction modes are initially separated, and their interaction is suddenly turned on. Since in our approach the dynamics of the entire system, including the conduction modes, is simulated, we also have information about the quantum state of bath. By analyzing the bath state we find signatures of the Kondo effect in the quench dynamics. We particularly discuss the dynamics of the entanglement along the MPS for different chain orderings. 
In \cref{sec:conclusion}, we summarize our results and draw our conclusions.

\section{Model and methods} \label{sec:model}
\subsection{Anderson Impurity Model}
Throughout this paper we consider the single-impurity Anderson model \cite{Anderson_PR61}, consisting of a single impurity which hybridizes with a free electron bath, modelling a half-filled conduction band:
\begin{equation}
\Ham_{\SIAM} = \Hloc + \Hcond + \Hhyb \;.
\end{equation}
The impurity site is described by the local Hamiltonian
\begin{equation}
\Hloc = \sum_{\sigma} \varepsilon_d \opddag{\sigma}\opd{\sigma} + U\, \opn{\uparrow}\opn{\downarrow} \;,
\end{equation}
where $\opddag{\sigma}$ creates an electron with spin $\sigma=\uparrow,\downarrow$ in the impurity orbital, 
$\opn{\sigma}= \opddag{\sigma}\opd{\sigma}$ is the number operator, and $U$ the on-site Hubbard repulsion. 
While our approach allows for arbitrary time-dependence in the local Hamiltonian, we restrict ourselves to a time-independent scenario 
for simplicity here. 
The conduction electrons are modelled as a half-filled free-electron bath:
\begin{align}
\Hcond = \sum_{\sigma}\sum_{k} \; \epsilon_k\, \opcdag{k\sigma} \, \opc{k\sigma}
\end{align}
where $\opcdag{k\sigma}$ creates an electron with energy $\epsilon_k$ in the conduction band, 
the kinetic energy $\epsilon_k$ being measured with respect to the chemical potential $\mu=0$. 
The impurity is coupled to the bath of free electrons through the hybridization term $\Hhyb$, allowing electrons to hop from the impurity 
into the bath and vice-versa:
\begin{equation} \label{eq:AIM_hybr_discrete}
\Hhyb = \sum_{\sigma} \sum_{k}  V_{k} \, \Big( \opddag{\sigma} \, \opc{k\sigma} + \opcdag{k\sigma} \, \opd{\sigma} \Big) \;.
\end{equation}
The hybridization matrix elements $V_k$, taken to be real, are connected to the imaginary part of the hybridization function 
self-energy $\Sigma_0(\omega)=\sum_k \frac{V_k^2}{\hbar\omega-\epsilon_k +i0^+}$. 

At zero temperature, the quantum state of such a closed system is pure, and hence can in principle be represented as a matrix product state. 
Finite temperatures, on the other hand, require us to use the density matrix formalism. 
Technically, this is possible through the matrix-product-operator approach, which, however, can be costly in practice, as it requires 
to prepare the thermal state through an imaginary-time evolution~\cite{PRL_Cirac_2004,PRL_Vidal_2004}. 

In this paper we follow the approach of Takahashi and Umezawa~\cite{ColPhen_Umezawa_1975}, representing the thermal density matrix 
$\rho_{\mathrm{cond}}\propto \exp(-\beta\Hcond)$ of the conduction electrons, with $\beta=1/k_BT$, as a pure state in a suitably 
enlarged Hilbert space. 
The idea is easy to understand. We consider the enlarged Hilbert space 
$\mathcal{H}_{\mathrm{cond}}\otimes\mathcal{H}_{\mathrm{anc}}$ built from the Hilbert space of the conduction electrons, 
$\mathcal{H}_{\mathrm{cond}}$, and an ancillary Hilbert space $\mathcal{H}_{\mathrm{anc}}$. 
We then prepare the enlarged system in a pure quantum state $|T\rangle$, represented using matrix product states, such that the partial trace over the ancillary modes yields the thermal density matrix for the conduction electrons, $\rho_{\mathrm{cond}}=\Tr_{\textrm{anc}}(|T\rangle\langle T|)$. 
%

\subsection{Thermofield transformation} 

We briefly summarize here the thermofield transformation. First, we rename the conduction modes, 
$\opc{k\sigma}\rightarrow \opc{1k\sigma}$, by adding the additional index '1'. 
We then add ancillary fermions, denoted by $\opc{2k\sigma}$, 
supplementing the conduction Hamiltonian with an ancillary bath term:
\begin{equation}\label{eq:Hcond_anc}
\Hcond=\sum_{\sigma}\sum_{k} \epsilon_k\, \Big( \opcdag{1k\sigma} \, \opc{1k\sigma } + \opc{2k\sigma} \, \opcdag{2k\sigma} \Big) \;.
\end{equation}
The ancillary fermions $\opc{2k\sigma}$ do not couple to either the impurity or the physical conduction electrons, and 
therefore will not affect the dynamics. 
Dropping spin indices for a while, we introduce two new fermionic operators as linear combinations of physical and ancillary fermionic operators, through the unitary thermofield transformation~\cite{ColPhen_Umezawa_1975, PRA_Vega_2015, PRL_Weichselbaum_2018, PRB_Plenio_2020}
\begin{equation} \label{eq:thermofield}
\left( \begin{array}{c} \opf{1k} \\ \opf{2k} \end{array} \right) =
\left( \begin{array}{rr}  \cos \theta_k  & - \sin \theta_k \\   \sin \theta_k & \cos \theta_k \end{array} \right) 
\left( \begin{array}{c} \opc{1k} \\ \opcdag{2k} \end{array} \right) \;,
\end{equation}
with its inverse given by
\begin{equation} \label{eq:thermofieldInverse}
\left( \begin{array}{c} \opc{1k} \\ \opcdag{2k} \end{array} \right) =
\left( \begin{array}{rr}  \cos \theta_k  & \sin \theta_k \\   -\sin \theta_k & \cos \theta_k \end{array} \right) 
\left( \begin{array}{c} \opf{1k} \\ \opf{2k} \end{array} \right) \;.
\end{equation}
Note that the transformation includes an additional particle-hole transformation on $\opf{2k}$  as compared to the original formulation \cite{ColPhen_Umezawa_1975,PRA_Vega_2015}, in order to maintain the particle number conversation of the Hamiltonian \cite{PRL_Weichselbaum_2018}.
Hence, in absence of the impurity, the thermal state is not represented by the vacuum state of $\opf{1k}$ and $\opf{2k}$, but rather by 
the vacuum $|\emptyset_1\rangle$ of $\opf{1k}$ and the fully occupied state $|\mathrm{F}_2\rangle$ of $\opf{2k}$, which in the following 
we will denote by $|\emptyset_1\rangle \otimes |\mathrm{F}_2\rangle = |\emptyset_1, \mathrm{F}_2\rangle$. 
Using \cref{eq:thermofieldInverse} we can show that the number operator $\opn{1k}=\opcdag{1k}\opc{1k}$ of the physical bath transforms as
\begin{align} \label{eq:thermofield_occ_op}
\begin{split}
\opcdag{1k}\opc{1k}&
=\cos^2(\theta_k)\opfdag{1k}\opf{1k} + \sin^2(\theta_k)\opfdag{2k}\opf{2k} \\
&+\cos(\theta_k)\sin(\theta_k)\left(\opfdag{1k}\opf{2k}+\opfdag{2k}\opf{1k}\right) \;.
\end{split}
\end{align}
Hence, the average physical electron occupation in the state $|\emptyset_1, \mathrm{F}_2\rangle$ is
\begin{align} \label{eq:thermofield_occ_expect}
\langle \emptyset_1, {\mathrm F}_2 |  \, \opcdag{1k} \, \opc{1k} | \emptyset_1, {\mathrm F}_2\rangle = \sin^2(\theta_k) \;.
\end{align}
We would like these occupations to follow the thermal distribution, given by the Fermi function $f_F(\epsilon)$. 
To this end, we make the choice \cite{PRB_Plenio_2020}
\begin{align} \label{eq:thermofield_angle}
	\sin^2(\theta_k)  \equiv  f_F(\epsilon_k) = 
	\frac{1}{\nep^{\beta\epsilon_k} + 1} \;.
\end{align}
Knowing how to prepare the thermal state in the basis of bath modes $\opf{1k}$ and $\opf{2k}$, we transform the Hamiltonian into 
this basis, by using the unitary transformation in \cref{eq:thermofieldInverse}. 
The hybridization term becomes
\begin{equation}
\sum_{k}  V_{k} \, \opddag{}\! \, \opc{1k}  = 
\sum_{k}  \Big( V_{1k} \, \opddag{}\! \, \opf{1k} + V_{2k} \, \opddag{}\! \, \opf{2k}  \Big) \;,
\end{equation}
with $V_{1k} = V_k \, \cos \theta_k$ and  $V_{2k} = V_k \, \sin \theta_k$. 
Originally coupled to the physical conduction electrons only, the impurity now interacts with both transformed modes, $\opf{1k}$ and $\opf{2k}$, with renormalized temperature-dependent couplings as visualized in \cref{fig:ChainCoefficients}(a,b). 
Hence, the temperature dependent thermofield transformation encodes finite temperature into the hybridization couplings, while the state in the basis of fermions $\opf{1k}$ and $\opf{2k}$ is independent of $T$. 
The conduction term including the ancillary bath, \cref{eq:Hcond_anc}, transforms as
\begin{equation}
\sum_{k} \epsilon_k\, \Big( \opcdag{1k} \, \opc{1k} + \opc{2k} \, \opcdag{2k} \Big)  = 
\sum_k \epsilon_k  \Big( \opfdag{1k} \, \opf{1k} + \opfdag{2k} \, \opf{2k} \Big) \;.
\nonumber
\end{equation}
When employing the thermofield method we need to simulate two independent baths of free fermions 
--- one being empty ($\opf{1k}$) and one being filled ($\opf{2k}$) ---, both interacting with the impurity only. 
While in principle a direct simulation in the star geometry --- using artificial long range interactions --- would be possible, we focus here on 
the chain geometry, following \cite{ArXiv_Kohn_2020}.
In particular, we apply two independent chain mappings for the empty and filled fermions $\opf{1k}$ and $\opf{2k}$, respectively. 
For the chain mapping there are mainly two options: 
1) The continuous bath can be discretized into a finite number of modes, e.g., by means of linear or logarithmic discretization, and mapped into a tight-biding chain using Lanczos' tridiagonalization algorithm~\cite{Wolf_PRB14,RMP_Bulla_2008,PRB_Wolf_2014_Cheb}. 
2) Employing the theory of orthogonal polynomials~\cite{ACM_Gautschi_1994}, a star-like bath can be transformed into a semi-infinite 
tight-binding chain~\cite{JMath_Chin_2010,PRL_Prior_2010}. 
This second approach can be particularly useful when working with structured hybridization functions. 
In this paper, we have used the orthogonal-polynomial-based mapping, since for an initial state with impurity decoupled from the conduction modes, it immediately allows us to work in the continuum limit. 

\subsection{Chain mapping with orthogonal polynomials}
First, let us turn to a continuum description of the baths. Denoting the half-bandwidth by $W$, we work with reduced dimensionless units $x=\epsilon/W$. 
In the continuum limit \cite{JPCM_Bulla_1997,RMP_Bulla_2008}, after reinstalling spin indices, we replace $\opf{ck\sigma}\rightarrow \opf{\chain\sigma}(x)$, 
with a Dirac delta anti-commutation relationship $\{\opf{\chain\sigma}(x),\opfdag{c'\sigma'}(x')\}=\delta_{cc'}\delta_{\sigma\sigma'}\delta(x-x')$, and recast the kinetic term as:
\begin{equation*}
	\Hcond = W \sum_\sigma \sum_{c=1}^2  \int_{-1}^{1} \! \ud x \; x \, \opfdag{\chain\sigma}(x) \, \opf{\chain\sigma}(x) \;,
\end{equation*}
and the hybridization term as:
\begin{equation*}
	\Hhyb = W \sum_{\sigma}  \sum_{c=1}^2  \int_{-1}^{1} \! \ud x \; V_{\chain}(x) \left(\opddag{\sigma}\, \opf{\chain\sigma}(x) + \Hc \right) \;.
\end{equation*}
%
Next, we carry out independent chain mappings for both chains. For that purpose, we define new fermionic operators as
\begin{align} \label{eq:orthPol_transform}
	\opa{\chain,n,\sigma} = \int_{-1}^{1} \! \ud x \; U_{\chain,n}(x) \, \opf{\chain\sigma}(x)\;,
\end{align}
with inverse transformation 
\begin{align} \label{eq:orthPol_transform_inv}
\opf{\chain\sigma}(x) = \sum_{n=0}^{\infty} U_{\chain,n}(x) \, \opa{\chain,n,\sigma} \;.
\end{align}
Here $U_{\chain,n}(x)=V_{\chain}(x) \, p_{\chain,n}(x)$ is a (real) unitary transformation provided the set of real polynomials $\{p_{\chain,n}\}$ are normalized 
and mutually orthogonal with respect to the corresponding weight function $V^2_{\chain}(x)$:
\begin{align}
	\int_{-1}^{1} \! \ud x \; V^2_{\chain}(x) \, p_{\chain,n}(x) \, p_{\chain,m}(x) = \delta_{n,m} \;.
\end{align}
This, in turn, implies that the fermionic operators defined by \cref{eq:orthPol_transform} satisfy the usual (anti-)commutation relation 
$\{\opa{\chain,n,\sigma},\opadag{\chain',n',\sigma'}\}=\delta_{\chain,\chain'}\delta_{\sigma,\sigma'}\delta_{n,n'}$.
Notice that the new creation (annihilation) operators are linear combinations of creation (annihilation) operators only. 
Hence, the empty (filled) bath state transforms into an empty (filled) chain, being a product state as well. 
This is the crucial advantage of the chain mapping introduced in Ref.\cite{ArXiv_Kohn_2020} as compared to the original $T=0$ chain mapping, 
where both empty and filled modes are transformed into a single chain, leading to an entangled state with partially filled chain sites. 
To carry out the transformation of the Hamiltonian we need the following property of orthogonal polynomials: The monic polynomials 
$\{\pi_{\chain,n}\}$, obtained by rescaling the normalized polynomials $\{p_{\chain,n}\}$ such that the coefficient of the leading degree term is one,
satisfy the recurrence relation \cite{ACM_Gautschi_1994,PRL_Prior_2010,JMath_Chin_2010,PRB_Schroeder_2016}
\begin{align}
	\pi_{\chain,n+1}(x) = (x-\alpha_{\chain,n})\pi_{\chain,n}(x)-\beta_{\chain,n}\pi_{\chain,n-1}(x) \;,
\end{align}
with recurrence coefficients $\{\alpha_{\chain,n}\}$ and $\{\beta_{\chain,n}\}$, uniquely defined by the weighting function $V^2_{\chain}(x)$. 
For the weighting function $V^2_{\chain}(x)$ with finite support $[a,b]$, it can be shown~\cite{JMath_Chin_2010} that 
these coefficients converge as $\alpha_{\chain,n}\rightarrow(a+b)/2$ and $\beta_{\chain,n}\rightarrow(b-a)^2/16$ for $n\rightarrow\infty$.
\begin{widetext}
Using the inverse chain mapping transformation, \cref{eq:orthPol_transform_inv}, and the recurrence relation, we transform the Hamiltonian into the new basis:
\begin{equation}
\Ham_{\AIM} = \Hloc  +  \sum_{\sigma} \sum_{c=1}^2 \Bigg( J_{\chain,0} \left(\opddag{\sigma} \, \opa{\chain,0,\sigma} + \Hc \right) 
+  \sum_{n=0}^{\infty}  \Big( E_{\chain,n} \, \opadag{\chain,n,\sigma}\opa{\chain,n,\sigma} + \big(J_{\chain,n+1} \opadag{\chain,n+1,\sigma}\opa{\chain,n,\sigma} + \Hc \big) \Big) \Bigg) \;.
\end{equation}
%
The chain coefficients are directly related to the recurrence coefficients of the orthogonal polynomials through
\begin{equation}
J_{\chain,0} = W\Big( \int_{-1}^{+1}\ud x \; V_{\chain}^2(x) \Big)^{\frac{1}{2}} \;, \hspace{10mm} 
J_{\chain,n\ge1} = W \sqrt{\beta_{\chain,n}} \;, \hspace{10mm} 
E_{\chain,n} = W\alpha_{\chain,n} \;.
\end{equation}
In practice, those coefficients are obtained numerically, using the routines of Refs.~\cite{ACM_Gautschi_1994,Gautschi_2004}.
\end{widetext}

\begin{figure}[t]
\centering
\includegraphics[width=8.6cm]{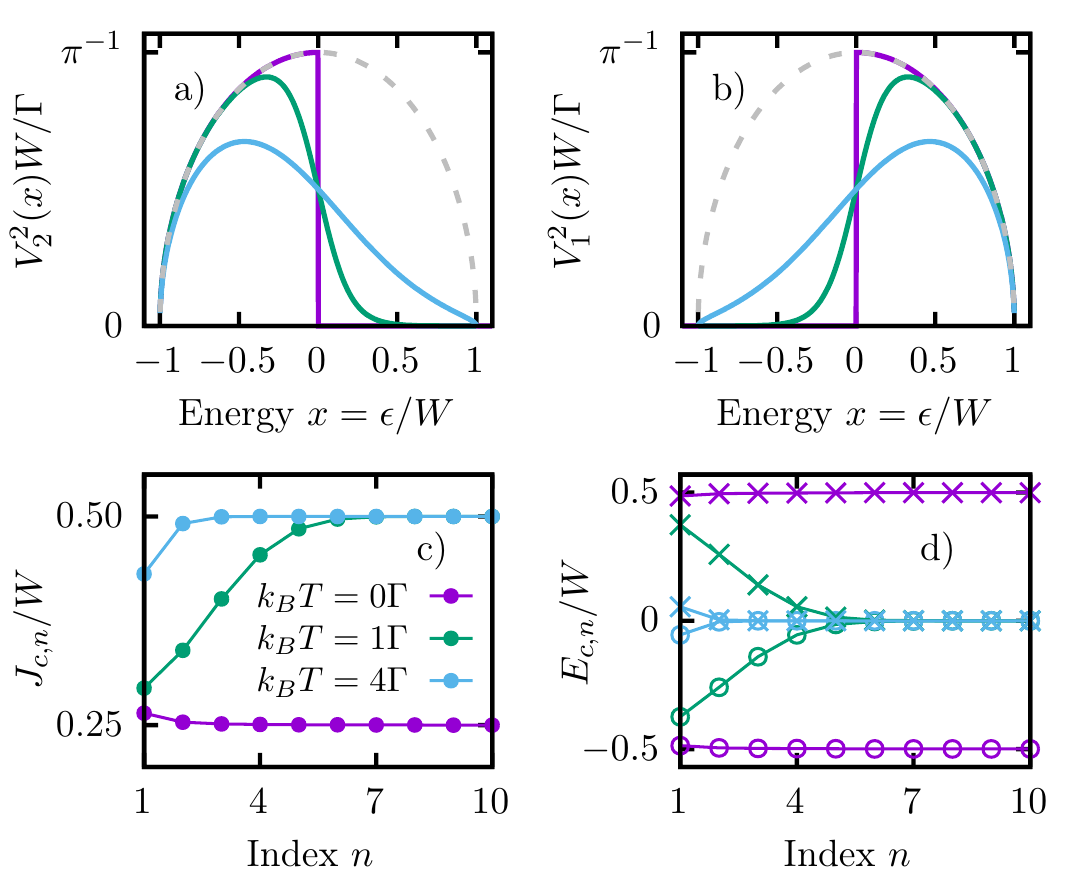}
\caption{(a,b) Renormalized hybridzation couplings $V_2^2(x)=V^2(x)\sin^2(\theta)$ (a) and  $V_1^2(x)=V^2(x)\cos^2(\theta)$ (b) for different temperatures $T$ (see legend panel (c)),
where $\sin^2(\theta)=f_F(x)$ is the Fermi function, to ensure the correct distribution of bath electrons, and $V^2(x)=\Gamma\sqrt{1-x^2}/\pi W$ (dashed line).
c) Couplings $J_{\chain,n}$ along the initially empty ($c=1$) and fully occupied ($c=2$) chains. 
d) On-site energies $E_{1,n}$ (crosses) and $E_{2,n}$ (open circles). 
Here, $J_{1,n}=J_{2,n}$ and $E_{2,n}=-E_{1,n}$ due to the symmetric hybridization function, $V(x)=V(-x)$.
}
\label{fig:ChainCoefficients}
\end{figure}
In \cref{fig:ChainCoefficients} we show the two renormalized couplings $V^2_{\chain}(x)$ in the continuum limit for the semi-circular hybridization $V^2(x)=\Gamma\sqrt{1-x^2}/\pi W$, and the corresponding chain coefficients $J_{\chain,n}$ and $E_{\chain,n}$. 
We clearly see that after only a few sites the chain coefficients converge towards the values expected from theory: At $T=0$ the renormalized hybridization functions have support $[0,1]$ ($V_1(x)$) and $[-1,0]$ ($V_2(x)$). 
Hence, the couplings converge as $J_{\chain,n}\rightarrow W/4$, while for on-site energies we find $E_{1,n}\rightarrow W/2$ and $E_{2,n}\rightarrow -W/2$. 
For $T>0$ instead, both $V_1(x)$ and $V_2(x)$ have support $[-1,1]$, implying $J_{\chain,n}\rightarrow W/2$ and $E_{\chain,n}\rightarrow 0$ for $n\rightarrow\infty$. 
For reasons of numerical convergence, it can be beneficial to truncate the support of $V_{\chain}(x)$ for the calculation of the chain coefficients, to eliminate regions where $V_{\chain}(x)$ falls below computational precision. 
This typically happens at low temperatures, where the Fermi function has a very small negligible tail.

\begin{figure}[t]
	\centering
	\includegraphics[width=8cm]{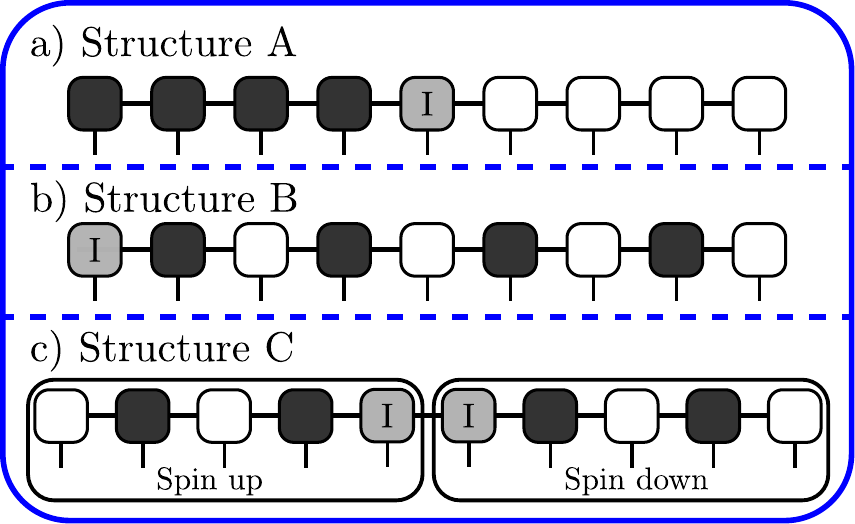}
	\caption{Different structures of the MPS: The impurity is visualized in light gray and empty (filled) chain sites are colored in white (black). 
	\textbf{A}) Structure suggested by the Hamiltonian, with spinfull fermionic sites (local dimension 4). Empty (left) and filled (right) chains are separated and connected to the impurity, placed in the middle of the MPS. The interaction is only nearest neighbor. 
	\textbf{B}) Interleaved ordering, with the impurity on the left and alternating filled and empty chain sites. In the MPS the interactions become next-nearest neighbor. 
	\textbf{C}) Same as \textbf{B}, with separated spin components. The local sites are spinless (local dimension two).}
	\label{fig:mpsorganization}
\end{figure}

To carry out simulations using MPS there is one more decision to make: How to order the chain sites in the MPS. 
This question is absolutely crucial for the simulation, since it affects the entanglement structure in the MPS, and therefore has major impact on the performance, as we will see. 
Three different possibilities have been considered in this paper:
\begin{itemize}
\item[\textbf{A)}] The most intuitive idea is to employ spinful fermionic sites, with the impurity placed in the middle of the MPS. The two chains, for empty and filled modes, are both connected to the impurity, one to the left and one to the right 
(see \cref{fig:mpsorganization}(a)). In this way, there are only nearest-neighbor interactions in the MPS, and each tensor represents both spin up and spin down states, with local (physical) dimension $d=4$, corresponding to states 
$|0\rangle, |\!\!\uparrow\rangle, |\!\!\downarrow\rangle, |\!\!\uparrow\downarrow\rangle$. 
This choice reflects the interaction structure of the Hamiltonian.
\item[\textbf{B)}] The second possibility is obtained by reordering the tensors of structure \textbf{A}. Here, the impurity is placed at the very first site of the MPS. 
The subsequent sites represent the two chains, with chain sites corresponding to the filled and empty chain, in an alternating fashion. Since the interaction within the two chains is nearest-neighbor, the interaction in the MPS is now up to
next-nearest neighbors. The impurity is interacting with the first site of the filled chain (second tensor in \cref{fig:mpsorganization}(b)) and the first site of the empty chain (third tensor in \cref{fig:mpsorganization}(b)). 
The idea behind this structure is the following. Imagine that during the dynamics an electron moves from the filled into the empty chain, creating an entangled particle-hole pair. 
%
Such a particle-hole pair will be travelling along the MPS without being much spatially separated in structure \textbf{B}, while a long-ranged entanglement is certainly required in structure \textbf{A}.  
\item[\textbf{C)}] The third structure follows the idea of structure \textbf{B}. However, instead of working with spinfull sites, we build the MPS with spinless sites, separating spin-up and spin-down degrees of freedom. 
This idea is suggested by the structure of the Hamiltonian: Spin-up and spin-down modes interact only at the impurity site.
It is well known that spatially separating the spins can be beneficial for numerical simulations \cite{PRB_Verstraete_2015,PRX_Bauernfeind_2017,PRL_Rams_2020}.
\end{itemize}

We carry out simulations using the 2-site version of the time-dependent variational principle (TDVP)~\cite{PRL_Haegeman_2011, PRB_Haegeman_2016, SIAM_Lubich_2015, AoP_Paeckel_2019}, which, in combination with the matrix product operator representation of the Hamiltonian, allows us to deal with next-nearest neighbor interaction. 
Using TDVP, it would also be possible to simulate more complicated networks \cite{PRB_Schroeder_2016, Scipost_Bauernfeind_2020, PRA_Kohn_2020}, which would be needed to split both spin degrees of freedom and empty/filled chains. Depending on the MPS ordering, we use bond dimensions between $D=150$ and $D=1600$ to reach convergence (see \cref{sec:bond_conv}) and a total truncated weight $w_t$ --- the summed probability of discarded states --- of $w_t=10^{-12}$ for the truncation of the MPS. 
We further employ a minimum bond dimension $D_{\min}\approx 10$, keeping even states with low probability, to reduce the projection error of TDVP (see \cref{sec:minD} for details). 
The time-step is fixed to be $\Delta t = 0.1 \hbar/W$.
We explicitly exploit the particle-number conservation of the Hamiltonian to speed up simulations. 
In practice, we must use a finite number of chain sites. 
We choose the number of sites such that no excitation --- being either a particle in the empty chain or a hole in the filled chain --- reaches the end of the chain (see also Ref.~\cite{PRB_Schroeder_2016}). 
For simulations up to time $t=5\hbar/\Gamma$, we typically use about 100 fermionic sites for each chain.

\section{Results} \label{sec:results}
We consider an impurity and conduction electrons initially decoupled from each other, and suddenly turn on, at time $t=0$, the hybridization coupling, evolving the system with a constant Hamiltonian $\Ham_{\AIM}$. 
More in detail, we initialize the system in the state $|\psi_0\rangle=|0\rangle\otimes|\emptyset_1, {\mathrm F}_2\rangle$, where $|0\rangle$ is the impurity vacuum and $|\emptyset_1, {\mathrm F}_2\rangle$ 
is the thermal state of the conduction electrons, represented as a pure state in the extended Hilbert space. For the hybridization of the impurity with the conduction electrons we choose a 
semi-circular form, $V^2(x)=\Gamma\sqrt{1-x^2}/\pi W$, where $x=\epsilon/W$ is the dimensionless energy, and $W$ half the bandwidth. 
Throughout this paper we fix the hybridization coupling $\Gamma$ such that $W=10\Gamma$. 
In the following, we study the dynamics of the combined system, with a special focus on the evolution of the entanglement entropy for the different MPS structures in \cref{fig:mpsorganization}.

\begin{figure}[t]
\centering
\includegraphics[width=8.5cm]{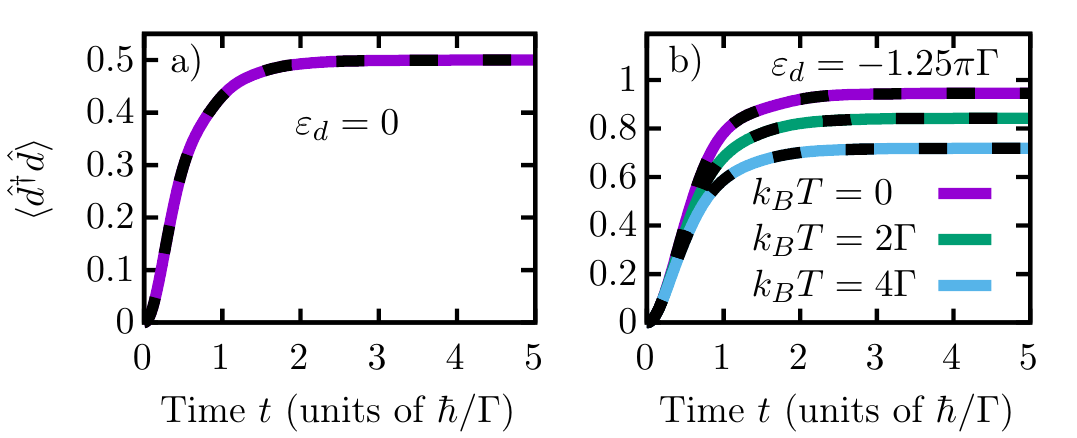}
\caption{(a,b) Dynamics of the $U=0$ impurity occupation $\langle \opddag{}\hat{d} \rangle$ for impurity energy level $\varepsilon_d=0$ (a) and $\varepsilon_d=-1.25\pi\Gamma$ (b) at different temperatures $T$. 
The dynamics is independent of $T$ for $\varepsilon_d=0$. 
Dashed lines represent results obtained from exact diagonalization, with linear discretization and 400 bath sites. MPS results were obtained using the structure \textbf{B} of \cref{fig:mpsorganization}. 
}
\label{fig:NonintImpurity}
\end{figure}

\begin{figure}[t]
	\centering
	\includegraphics[width=8.5cm]{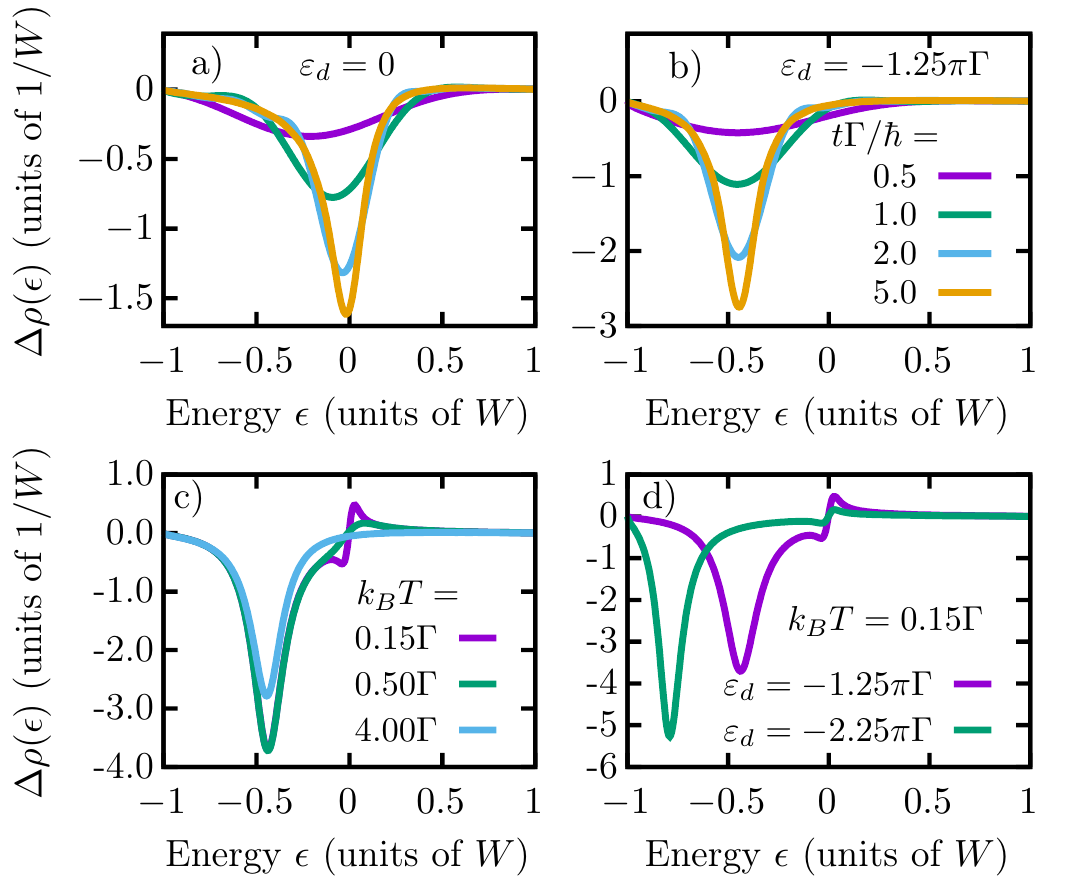}
	\caption{(a,b) Conduction band occupation density $\Delta\rho(\epsilon)$ at different times $t$,  for
		$\varepsilon_d=0$ (a) and $\varepsilon_d=-1.25\pi\Gamma$ (b) at temperature $k_BT=4\Gamma$. A well defined negative peak close to the impurity energy level appears. 
		(c+d) Converged $\Delta\rho(\epsilon)$ in the long-time limit for different temperatures $T$ (c) and $\varepsilon_d$ (d). At low temperatures a kink at the Fermi energy $\epsilon_f$ emerges (c), which is reduced as the impurity energy level moves away from $\epsilon_f$ (d).
	}
	\label{fig:NonintBathOccupations}
\end{figure}

\subsection{Noninteracting case $U=0$}
For $U=0$ the Anderson model reduces to the resonant level model, where spin degrees of freedom are decoupled. 
Hence, we can forget about the spin index and consider spinless fermions instead. 
In this case structures \textbf{A} and \textbf{B} in \cref{fig:mpsorganization} are simplified, and we use fermionic sites with local dimension $d=2$. 
Structure \textbf{C} 
will not be considered in this section.

\paragraph{Impurity occupation.}
First, let us discuss the quench dynamics of impurity occupation. 
At $U=0$, we compare our results obtained from the MPS approach using structure \textbf{B} with exact diagonalization (ED) results, finding perfect agreement between the two methods.
Since the impurity is initially empty, it starts to fill up at $t\ge0$, as shown in \cref{fig:NonintImpurity}.  
For $\varepsilon_d=0$, \cref{fig:NonintImpurity}(a), we observe a very smooth convergence towards $\langle \opddag{}\opd{}\! \rangle = 1/2$, a value consistent with the particle-hole symmetry of the final Hamiltonian. 
In this case, moreover, the dynamics is independent of temperature. 
For $\varepsilon_d=-1.25\pi\Gamma$, instead, as the impurity level lies below the Fermi energy $\epsilon_f=0$,  the impurity occupation converges towards a temperature-dependent 
steady-state value $\langle \opddag{}\opd{}\rangle \! \!> 1/2$, see \cref{fig:NonintImpurity}(b).
As expected, the equilibrium occupation goes towards $\langle \opddag{}\opd{} \!\rangle =1/2$ as temperature increases. 

\paragraph{Conduction electron density.}
%
We have seen that the initially empty impurity is populated during the dynamics. Particle number conservation implies that the conduction modes loose exactly the number of electrons that is gained by the impurity. Our method computes the dynamics of the entire system, including the conduction modes. Hence, we are able to study also the dynamics of the bath. 
In this paper we focus on the occupation of the conduction modes, although other quantities might in be calculated as well. 
The quantity we will calculate is the time-dependent expectation value 
\begin{equation} \label{eq:res_bath_occ}
\Delta\rho_{k}(t) =  \langle \psi(t) | :\! \opcdag{1k}\opc{1k}\! : | \psi(t) \rangle \;. 
\end{equation}  
of the conduction electron occupation number operator  
\[ 
:\! \opcdag{1k}\opc{1k}\! :\; \stackrel{\mathrm{def}}{=} \; \opcdag{1k}\opc{1k} - 
\langle \psi_0 | \opcdag{1k}\opc{1k} | \psi_0\rangle \;. 
\] 
Here, subtracting the initial state value is a device, akin to normal ordering, which takes care of the infinite number of electrons in the bath, and captures only the change in conduction electron density induced by the hybridization.
%
%

Details on the practical evaluation of this expression are given in \cref{sec_app:bath_occ}.
In the continuum limit, we calculate  $\Delta\rho(x,t)$ in proper energy units, $\epsilon=Wx$. 
We show it in \cref{fig:NonintBathOccupations}(a+b), for a temperature $k_BT=4\Gamma$.
Starting from $\Delta\rho(x,t=0)\equiv0$, we observe the growing of a peak close to the impurity level energy $\varepsilon_d$, similarly to what has been found for the Spin-Boson model ~\cite{PRB_Schroeder_2016}. 
Note that $\Delta\rho$ is predominantly negative, since particle conservation requires 
\[ \langle \psi(t) |\opddag{}\opd{}\!\! |\psi(t)\rangle = -\int_{-1}^{+1}\!\! \ud x  \; \Delta\rho(x,t) \] 
at any time. As temperature is reduced, \cref{fig:NonintBathOccupations}(c), we observe the appearance of a kink at the Fermi energy $\epsilon_f$, which we easily understand in the limit $T\rightarrow 0$: The conduction bath is completely filled below the Fermi energy, and empty above. The tail of the spectral weight -- corresponding to the local impurity level -- drains some of the initially occupied modes below $\epsilon_f$, and provokes the occupation of some initially empty modes above $\epsilon_f$. As we move the impurity level further away from the Fermi energy, and by that lowing the spectral weight at $\epsilon_f$, the kink is clearly reduced in size, \cref{fig:NonintBathOccupations}(d).
Once again, we benchmarked our calcualtions through comparison with ED data (not shown). 
As we will show later on, $\Delta\rho(x,t)$ can even contain information about many-body physics, in particular the Kondo effect.

\begin{figure}[t]
\centering
\includegraphics[width=8.5cm]{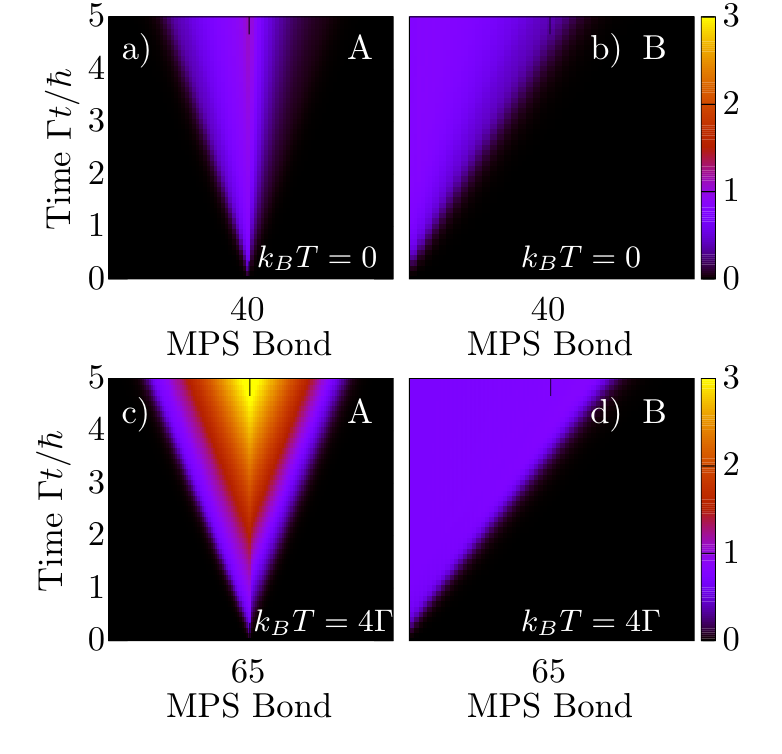}
\includegraphics[width=8.5cm]{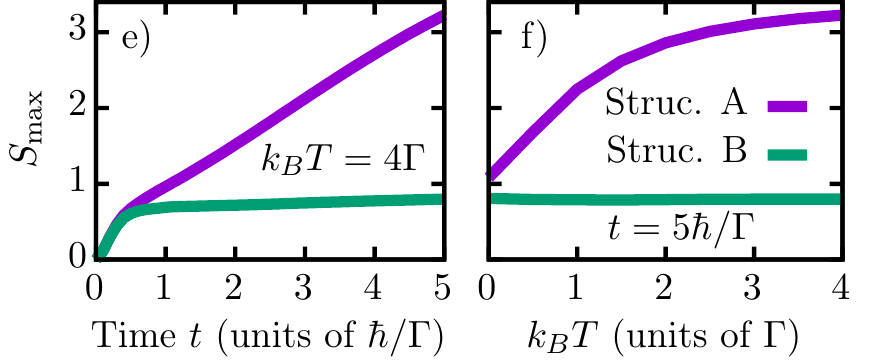}
\caption{Color plots: Dynamics of the entanglement entropy within the MPS in the noninteracting case $U=0$. Structures \textbf{A} (a+c) and \textbf{B} (b+d) are considered, with conduction electrons at temperature $T=0$ (a+b) and $T=4\Gamma$ (c+d). (e+f) Maximum entanglement entropy $S_{\text{max}}$ along the MPS as a function of time at fixed temperature $k_BT=4\Gamma$ (e) and as a function of temperature at fixed time $t=5\hbar/\Gamma$ (f). 
Structure \textbf{A} shows significantly stronger entanglement growth as temperature increases, while structure \textbf{B} has almost no entanglement growth and is independent of temperature.
}
\label{fig:NonintEnt}
\end{figure}

\paragraph{Entanglement.}
We now turn to the analysis of the entanglement dynamics. 
To quantify the amount of entanglement we calculate the entanglement entropy $S_l$ between the first $l$ sites of the MPS and the rest of the system. 
Since our initial state $|\psi_0\rangle=|0\rangle\otimes|\emptyset_1, {\mathrm F}_2\rangle$ --- an empty impurity and the bath in the thermal state --- is represented by a product state, the entanglement is zero along the MPS for $t=0$. 
For $t>0$, excitations --- particles in the empty chain or holes in the filled chain --- are created in the vicinity of the impurity. 
Hence, we observe the entanglement to grow (see \cref{fig:NonintEnt}) starting from the impurity's position in the MPS. Notice that the impurity is placed in the middle of the MPS for 
structure \textbf{A} and on the left in structure \textbf{B}. 
The region of nonzero entanglement is growing during the dynamics in a light-cone-like fashion, due to the spreading of excitations along the chains. 
We note a slight asymmetry in the entanglement of structure \textbf{A}, due to the initial state: Since we start with an empty impurity, and particle number is conserved, 
only the filled chain is able to interact with the impurity at $t=0$, leading to an initial entanglement predominantly between impurity and filled chain. 
Overall, we find the entanglement's magnitude to be similar for both MPS structures at $T=0$.

At higher temperature, $k_BT=4\Gamma$ (see \cref{fig:NonintEnt}(c,d)), we note that entanglement is spreading faster. Hence, longer chains are needed, independently of the MPS structure. However, the most striking effect of a higher temperature is the significant increase of entanglement in structure \textbf{A}: 
While structure \textbf{B} shows similar entanglement as for $T=0$, we observe a massive increase in structure \textbf{A}, mostly in the middle of the MPS,
indicating a strongly increasing entanglement between the empty and filled chain.

\begin{figure}[t]
\centering
\includegraphics[width=8.5cm]{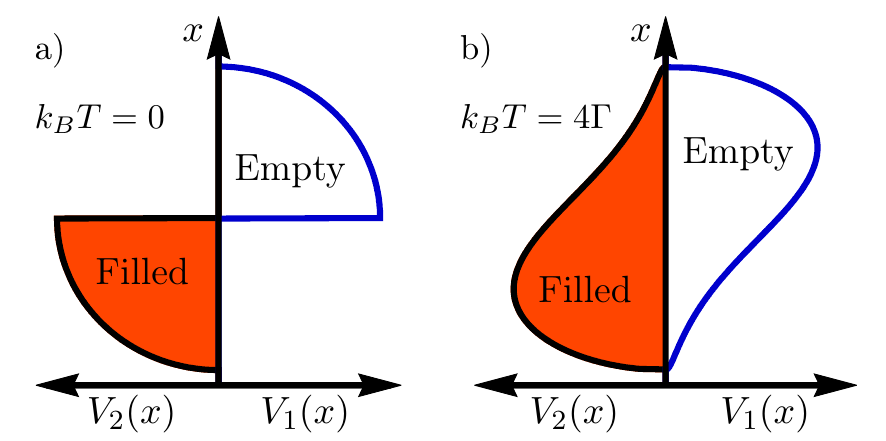}
\caption{(a,b): Effective hybridization functions $V_2(x)$ and $V_1(x)$ for the filled and empty bath, respectively, at temperatures $k_BT=0$ (a) and $k_BT=4\Gamma$ (b). 
$x=\epsilon/W$ is the dimensionless energy. At zero temperature, the effective hybridization functions do not overlap, touching only at the Fermi energy ($x=0$); For $T>0$ they do have a nonzero overlap, allowing particles to travel from the filled to the empty bath. 
}
\label{fig:hybOverlap}
\end{figure}

\begin{figure}[t]
	\centering
	\includegraphics[width=8.5cm]{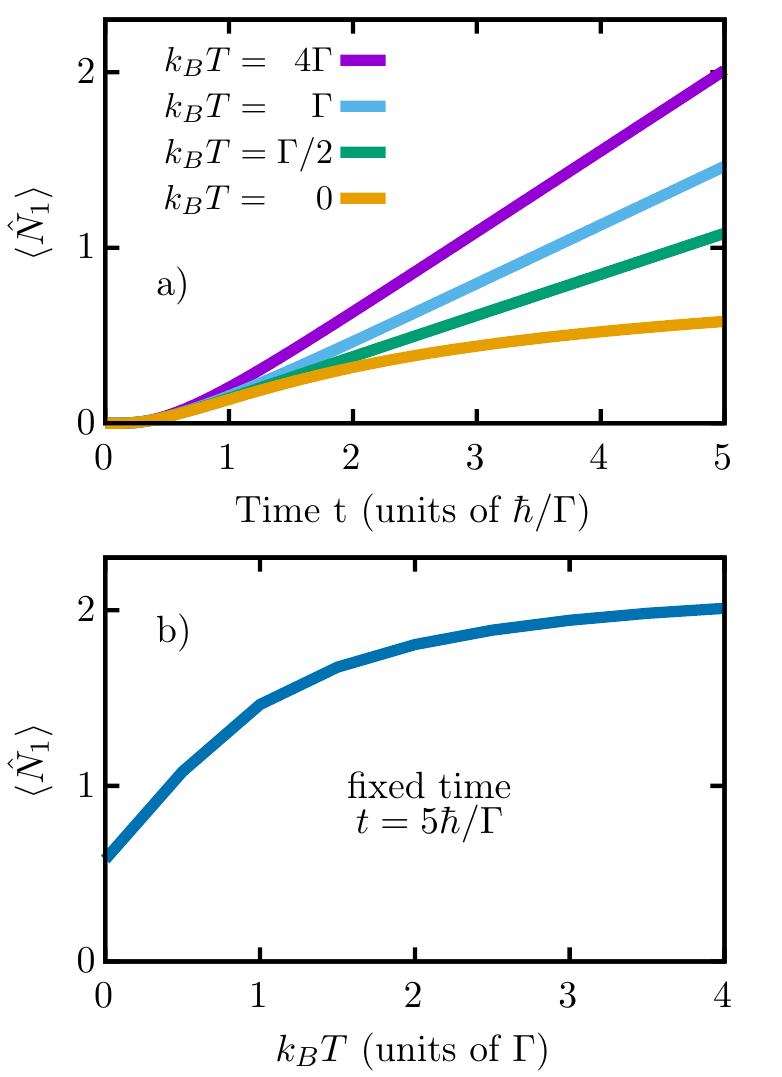}
	\caption{(a,b): Total number of electrons in the initially empty chain $\langle \psi(t) | \hat{N}_{1} |\psi(t) \rangle$ as a function of time for increasing temperature $k_BT$ (a), and at  
	fixed time $t=5\hbar/\Gamma$ as a function of temperature $k_BT$ (b).
	Notice that $\langle \hat{N}_{1} \rangle$ is independent of the MPS structure.
	}
	\label{fig:n1_chainOccupation}
\end{figure}

We define the maximum entanglement entropy $S_{\text{max}}$ along the MPS as $S_{\text{max}}=\max_{l} S_l$. 
We find (see \cref{fig:NonintEnt}(e)) that $S_{\text{max}}$ for $k_BT=4\Gamma$ linearly increases for structure \textbf{A}, while, after some initial increase, it stays almost constant for structure \textbf{B}. 
A linearly increasing entanglement entropy requires the bond dimension to grow exponentially in time, and thus strongly limits the accessible simulation times. 
As shown in \cref{fig:NonintEnt}(f), the entanglement highly depends on temperature for the structure \textbf{A}. 
These  observations are in agreement with the findings of Ref.~\cite{PRB_Millis_2017}. 
At $T=0$ the effective hybridization functions $V_1^2(x)$ and $V_2^2(x)$ only touch at the Fermi energy $x=0$ (see \cref{fig:hybOverlap}(a)). 
In this case, Ref.~\cite{PRB_Millis_2017} found the entanglement entropy to grow only logarithmically. 
At $T>0$, instead, the hybridization functions do overlap on a finite interval, see \cref{fig:hybOverlap}(b), leading to linear entanglement growth~\cite{PRB_Millis_2017}. 
Remarkably, merging the two chains, as we suggested in structure \textbf{B},  \cref{fig:mpsorganization}(b), results in a temperature-independent maximum entanglement (see \cref{fig:NonintEnt}(f)).

To add further intuition for this behavior, we measure the total number of particles in the initially empty chain, through the corresponding number operator (written here for spinless fermions, for simplicity) average:
\begin{equation}
\langle \psi(t) | \hat{N}_{1} |\psi(t) \rangle = \sum_{n=0}^{\infty} 
\langle \psi(t) | \opadag{1,n}\opa{1,n} |\psi(t) \rangle \;.
\end{equation}
$\langle \psi(t) | \hat{N}_{1} |\psi(t) \rangle$ counts how many particles flow from the filled chain  --- after passing through the impurity --- into the empty chain ``1''. 
\cref{fig:n1_chainOccupation}(a) shows that $\langle \hat{N}_{1}  \rangle$ increase linearly in time at finite temperature and sub-linearly at $k_BT=0$, similar to the entanglement in Ref.~\cite{PRB_Millis_2017}. \cref{fig:n1_chainOccupation}(b) shows that the temperature dependence of $\langle \hat{N}_{1} \rangle$ at fixed time $t=5\hbar/\Gamma$ agrees qualitatively well with our findings for the entanglement, see \cref{fig:NonintEnt}(f). 
Notice that any particle leaving the filled chain ``2'' creates a hole there. Hence, the dynamics creates particle-hole pairs: particles created in the empty chain ``1'' and holes in the filled chain ``2''. 
Our results suggest that such particle-hole pairs carry the entanglement, leading to an overall entanglement growth between the two chains. 

\begin{figure}[t]
\centering
\includegraphics[width=8.5cm]{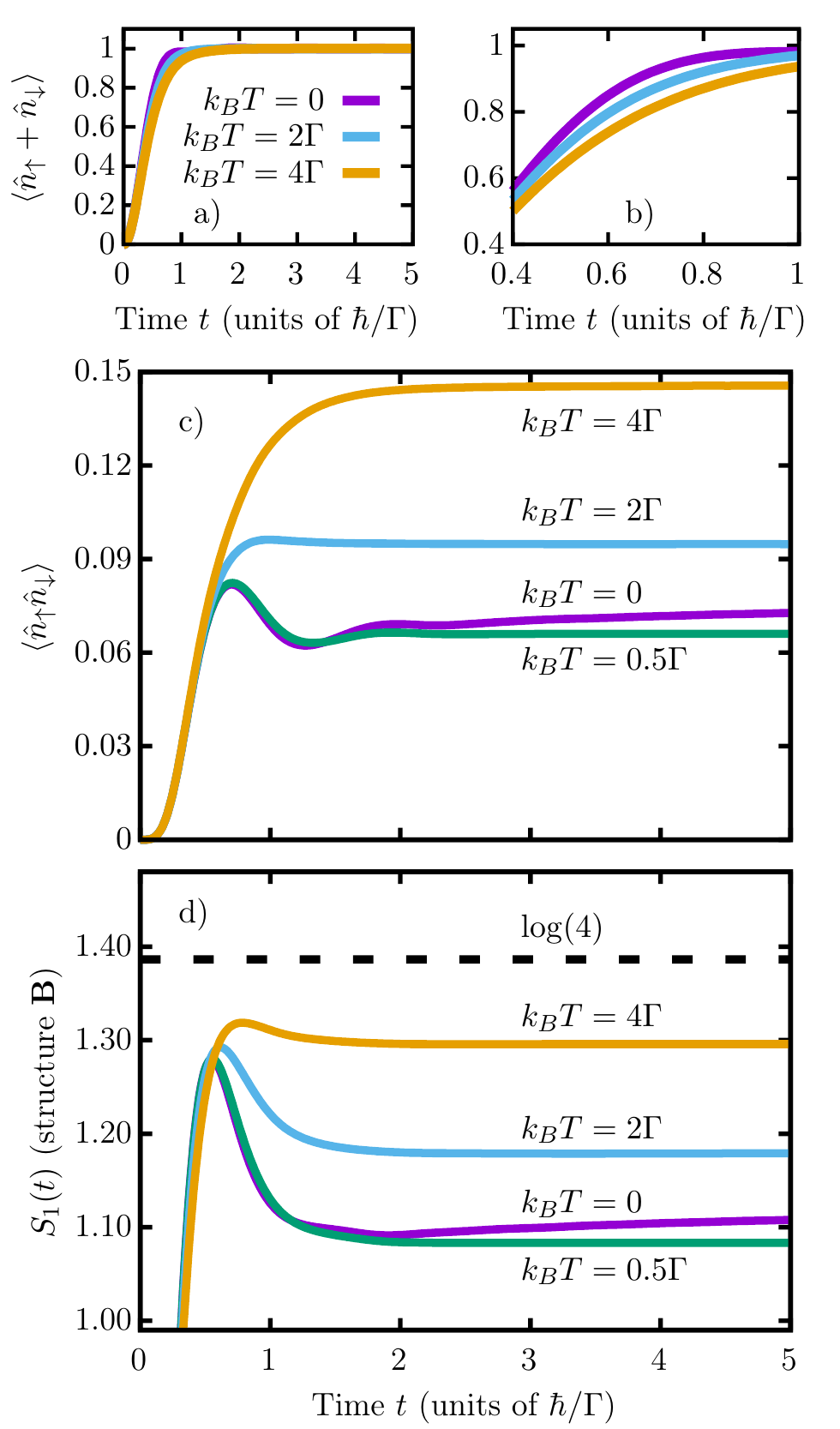}
\caption{(a,b) Dynamics of the total impurity occupation 
$\langle \hat{n}_{\uparrow} + \hat{n}_{\downarrow} \rangle$ 
for $U=2.5\pi\Gamma$ for the particle-hole symmetric choice $\varepsilon_d=-U/2$. 
(c) Double occupancy $\langle \hat{n}_{\uparrow} \hat{n}_{\downarrow} \rangle$.
(d) Entanglement entropy between impurity and bath, $S_1$. 
Dashed line in (d) indicates the theoretically maximum value of the entanglement entropy, $S_1=\log(4)$.
Data are obtained the using MPS ordering structure \textbf{B} (see \cref{fig:mpsorganization}).
}
\label{fig:IntImpurity}
\end{figure}

\subsection{Interacting case $U>0$}
Turning on the on-site interaction, we need to consider spinfull fermions, with spin-up and spin-down electrons in the impurity interacting through Coulomb repulsion $U$. 
Let us start with a brief analysis of the dynamics, with fixed interaction $U=2.5\pi\Gamma$ and energy level $\varepsilon_d=-1.25\pi\Gamma$, where the model is particle-hole symmetric ($U=-2\varepsilon_d$) with an estimated Kondo temperature $k_bT_K=0.07\Gamma$.
Hence, the impurity occupation --- starting again from zero --- converges towards $\langle \opddag{\sigma}\opd{\sigma}\rangle \rightarrow 1/2$ for both spin-up and spin-down at any temperature, with total impurity occupation $\langle \opn{\uparrow}+\opn{\downarrow}\rangle \rightarrow 1$ (see \cref{fig:IntImpurity}(a)). 
In contrast to the non-interacting case, however, the dynamics of the impurity occupation does show some small temperature dependence before reaching convergence, as visualized by an appropriate zoom-in, see \cref{fig:IntImpurity}(b). 
The double occupancy, shown in \cref{fig:IntImpurity}(c), is equivalent to the probability to find the impurity in the filled state $|\!\!\uparrow\downarrow\rangle$, and shows a clear (non-monotonic) temperature dependence in its final value. 
Note that the curve for $k_BT=0$ converges much slower than the remaining ones. We believe that the slow convergence --- and, connected to that, also the nonmonotonicity in temperature --- is related to the building up of the Kondo effect, for which slow convergence of the Greens function has been observed previously at low temperatures \cite{ArXiv_Kohn_2020}(supplementary material). 
Similar behavior is found for the entanglement entropy between the impurity and the free electron bath, including the non-monotonic temperature dependence and the slow convergence for $T=0$. Notice that, as temperature gets higher, the entanglement entropy tends towards its maximium possible value $S_1=\log(4)$.

\begin{figure}[t]
\centering
\includegraphics[width=8.5cm]{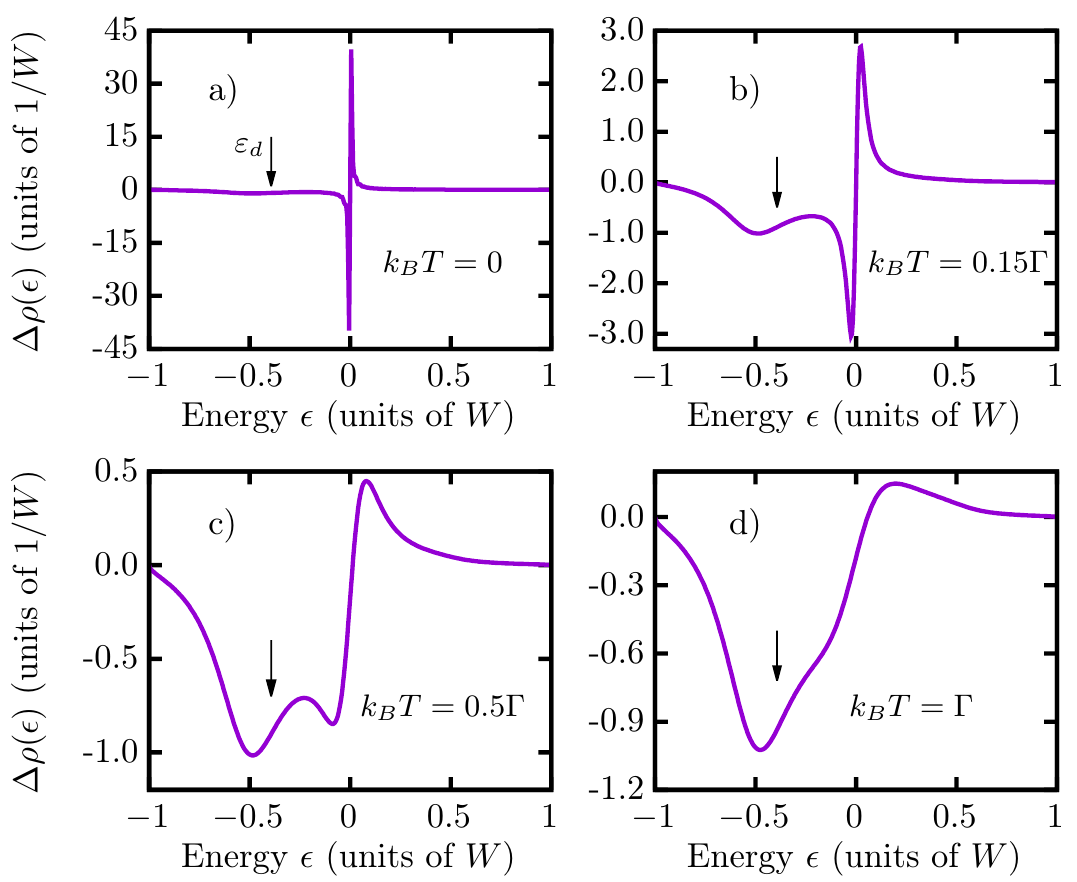}
\caption{Conduction electron occupation density $\Delta\rho(\epsilon,t)$ of spin-up conduction modes (spin-down is equivalent) at temperatures $T=0$ (a), $k_BT=0.15\Gamma$ (b), $k_BT=0.5\Gamma$ (c) and $k_BT=\Gamma$ (d). Curves are taken at times $t=30\hbar/\Gamma$ ($k_BT=0$), $t=15\hbar$/$\Gamma$ ($k_BT=0.15\Gamma$), $t=10\hbar$/$\Gamma$ ($k_BT=0.5\Gamma$) and $t=5\hbar$/$\Gamma$ ($k_BT=\Gamma$). For $T>0$ they are converged with respect to time and do not change anymore. Peaks are observed close to the impurity level $\varepsilon_d=-1.25\pi\Gamma$, marked through arrows, and for low temperatures around the Fermi energy, indicating the presence of the Kondo effect. The Kondo temperature is $k_BT_K=0.07\Gamma$.}
\label{fig:IntOccup}
\end{figure}

\begin{figure*}[t]
\centering
\includegraphics[width=16cm]{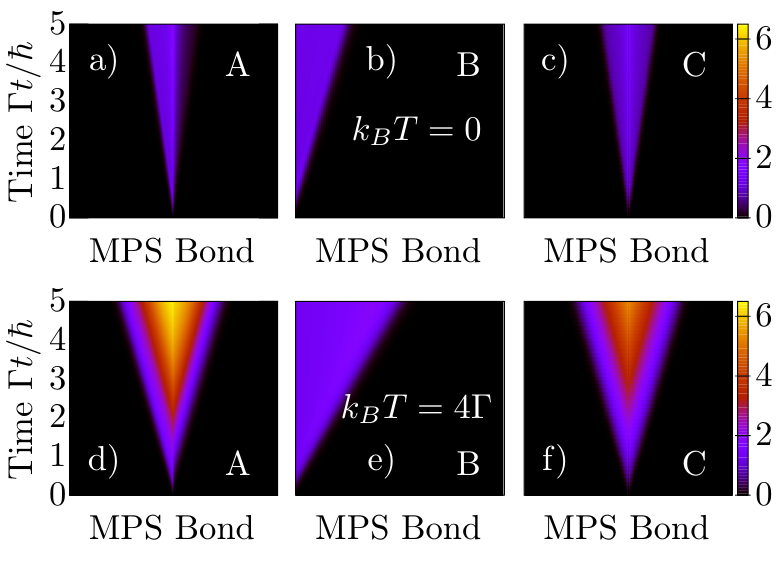}\\
\includegraphics[width=16cm]{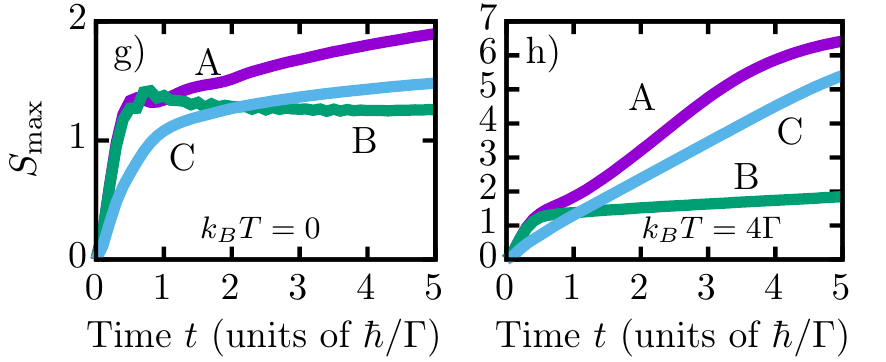}
\caption{Entanglement dynamics along the MPS at temperatures $k_BT=0$ (a-c) and $k_BT=4\Gamma$ (d-f), for structures \textbf{A}, \textbf{B}, and \textbf{C}, as illustrated in \cref{fig:mpsorganization}, at fixed interaction $U=2.5\pi\Gamma$, for the particle-hole symmetric case $\varepsilon_d=-U/2$. 
Bottom: Dynamics of the maximum entanglement entropy $S_{\text{max} }$ at fixed $k_BT=4\Gamma$ (g) and temperature dependence of $S_{\text{max} }$ at the final simulation time $t=5\hbar/\Gamma$ (h). In panel (h) the curve for structure \textbf{A} is a lower bound for $S_{\text{max}}$, as convergence with respect to the bond dimension has not been reached (see \cref{sec:bond_conv}). Filled and empty chains are made of 90 sites each.
}
\label{fig:IntEntanglement}
\end{figure*}

\paragraph{Conduction electron density.}
In the non-interacting case, we saw that the conduction electron occupation density $\Delta\rho(\epsilon,t)$ develops a (negative) peak around the impurity energy level at $\varepsilon_d$. 
In the interacting case, we find signatures of the Kondo peak at low temperatures. 
The Kondo effect manifests itself through two peaks of opposite sign around the Fermi energy $\epsilon_f=0$, which are similar to the kink at $U=0$, but significantly more pronounced.
Hence, the increased impurity spectral weight around the Fermi energy -- due to the formation of the Kondo cloud singlet -- results in two peaks of different sign in $\Delta\rho(\epsilon,t\to \infty)$.
As we increase $T$ above the Kondo temperature $k_BT_{K}=0.07\Gamma$, the two peaks close to $\epsilon_f$ disappear, just like the Kondo peak in the impurity spectral function, leaving peaks corresponding to the impurity level. 
Note that at $U\ne0$ there is a second impurity level at $\varepsilon_d+U$, corresponding to a fully occupied impurity. However, this state is not probed in our scenario, since we are starting from an empty impurity, and the doubly occupied state has little impact on the dynamics. However, we can probe this level starting from the fully occupied impurity (see \cref{sec:startFilled} for details). 
It is worth to mention that the convergence of $\Delta\rho(\epsilon,t)$ with respect to time is strongly temperature dependent, with faster convergence for higher temperatures, again, similarly to the convergence of the Green's function~\cite{ArXiv_Kohn_2020}. At $T=0$ we did not even reach convergence at time $t=30\hbar/\Gamma$, where peaks at $\epsilon=0$ are still growing. 

\paragraph{Entanglement.}
Turning to the entanglement, we investigate all MPS orderings illustrated in \cref{fig:mpsorganization}, where, additionally to structures \textbf{A} and \textbf{B} with spinfull sites, we consider structure \textbf{C} with spatially separated spin-up and spin-down. 
For $T=0$, we find similarly low entanglement for all structures, as shown in \cref{fig:IntEntanglement}(a-c)

Once again, for $k_bT=4\Gamma$, we observe the entanglement to grow strongly for structure \textbf{A}, with linearly increasing maximum entropy. While for our mixed structure \textbf{B} entanglement stays significantly lower (see\cref{fig:IntEntanglement}), we also identify a slight linear increase here. For structure \textbf{C}, where spin-up sites are separated from spin-down sites, the entanglement structure is similar to the one of structure \textbf{A}, with massive entanglement growth during the dynamics. Interestingly, the blow up of entanglement is observed in the middle of the MPS, suggesting that spin-up and spin-down sites are getting heavily entangled at higher temperature. Hence, the separation of spins unavoidably leads to strong entanglement growth, independent of the explicit structure used to represent the bath. This might affect all finite temperature generalizations of approaches exploiting a spatial separation, like the one recently developed by Rams \textit{et al.} to simulate transport through an impurity\cite{PRL_Rams_2020}.

\begin{figure}[t]
\centering
\includegraphics[width=8.5cm]{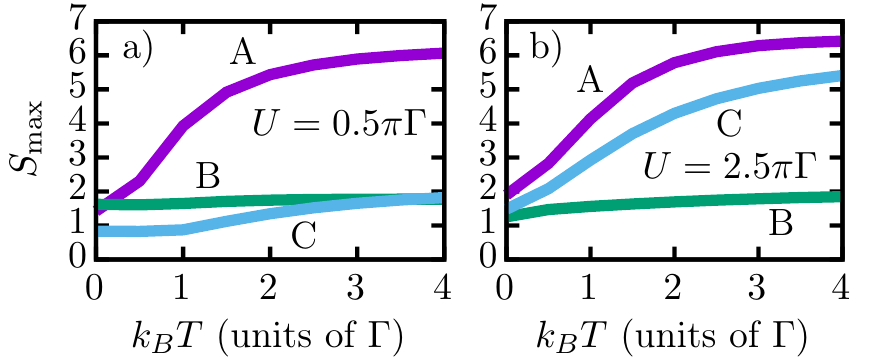}
\includegraphics[width=8.5cm]{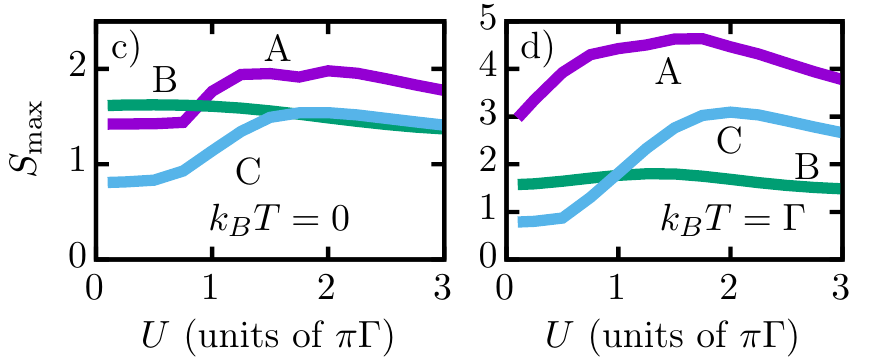}
\caption{Maximum entanglement entropy $S_{\mathrm{max}}$ along the MPS as a function of temperature $k_BT$ (a,b) and interaction $U$ (c,d), at fixed time $t=5\hbar/\Gamma$ for MPS structures \textbf{A}, \textbf{B}, and \textbf{C} (see \cref{fig:mpsorganization}). In panels (a,b) we fix the interaction to $U=0.5\pi\Gamma$ (a) and $U=2.5\pi\Gamma$ (b), while in panels (c,d) temperatures $k_BT=0$ (c) and $k_BT=\Gamma$ (d) are kept constant. Here $\epsilon_d=-1.25 \pi \Gamma$.}
\label{fig:Entanglement_difU}
\end{figure}

Let us now investigate the effect of different physical model parameters. To simplify the discussion we focus on the maximum entanglement entropy encountered during the dynamics up to time $t=5\hbar/\Gamma$, while the general entanglement structure along the MPS was observed to be similar to that discussed previously. 
\cref{fig:Entanglement_difU} shows the temperature dependence of the maximum entanglement entropy, for $U=0.5\pi\Gamma$ (a) and the particle-hole symmetric choice $U=2.5\pi\Gamma$ (b).
For both values of $U$, we observe a strong temperature dependence for structure \textbf{A}, while structure \textbf{B} is significantly less sensitive to temperature, similarly to what we found in the noninteracting case. 
Structure \textbf{C}, instead, shows little $T$-dependence for small interactions, but strong dependence for $U=2.5\pi\Gamma$. 
This behavior is easy to understand: We have seen in \cref{fig:IntEntanglement} that the entanglement for $U=2.5\pi\Gamma$ and $k_BT=4\Gamma$ grows strongly in between spin-up and spin-down modes. At $U=0$, however, spin-up and down are totally decoupled. Therefore, at low values of $U$ the entanglement growth between the opposite spins is still modest. 

To analyze the effect of the interaction in more detail, we show the maximum entanglement entropy as a function of $U$ in the bottom row of \cref{fig:Entanglement_difU}. Structures \textbf{A} and \textbf{C} display a non-monotonic behavior with maximum entanglement in the order of $U\approx (1\div 2)\Gamma$, while the spinfull interleaved ordering \textbf{B} is rather independent from the interaction. 
Hence, from an entanglement point of view, structure \textbf{A} never seems advantageous. The interleaved ordering with spin splitting, structure \textbf{C}, shows low entanglement at weak coupling and at low temperatures. In several scenarios, including high temperature at intermediate interactions, however, the entanglement grows strongly for orderings \textbf{A} and \textbf{C}, and structure \textbf{B} is able to capture the dynamics much more efficiently. 

\section{Conclusions} \label{sec:conclusion}
%
%
We have studied the dynamics of the quenched Anderson model in a wide range of temperatures $T$ and interactions $U$. Employing the chain geometry for the two conduction electron baths emerging from the thermofield approach, we have shown that the entanglement can massively depend on the ordering of the chain sites in the MPS. 

While at zero temperature all orderings considered here show slow-to-modest entanglement growth, the situation changes dramatically at higher temperature: It strongly grows if either the empty/filled chains or sites with different spin are spatially separated. 
Instead, merging the chains with alternating empty and filled sites --- such that the interaction terms in the MPS are next-nearest neighbor --- leads to significantly lower entanglement growth, allowing for much longer simulations with low numerical resources. 
For the separation of filled/empty chains we have reasoned that the growing entanglement is due to the increased rate at which particle-hole pairs are created, following from the overlap of effective hybridization functions at finite temperature. 
Our analysis has shown that, in non-equilibrium situations, it is not necessarily beneficial to mimic the Hamiltonian structure in the MPS. 
Instead, the ongoing physical processes, such as the movement of particles, determines the entanglement properties. 

Furthermore, we have shown that the analysis of the conduction bath --- available when simulating the full dynamics of system and ``environment'' --- can reveal interesting many-body physics, like the Kondo-effect. 
As an outlook for further research, it would be interesting how the star geometry would perform at finite temperature. 
Our results imply that the separation of filled and empty baths would lead to strong entanglement growth also in the star-geometry. 
However, a mixed ordering according to the energy of the modes might be a low-entanglement candidate for the star geometry.

\section*{Acknowledgements}
Research was partly supported by EU Horizon 2020 under ERC-ULTRADISS, Grant Agreement No. 834402.
GES acknowledges that his research has been conducted within the framework of the Trieste Institute for Theoretical Quantum Technologies (TQT).
Simulations were performed using the ITensor library \cite{itensor}.
\appendix

\section{Calculating the conduction occupation density} \label{sec_app:bath_occ}
In this appendix we provide details for the calculation of the residual bath occupation density 
$\Delta\rho(x,t)$. The starting point is the transformation of 
$:\! \opcdag{1k}\opc{1k}\! := \opcdag{1k}\opc{1k} - \langle \psi_0| \opcdag{1k}\opc{1k}|\psi_0\rangle$,  see Eq.~\eqref{eq:thermofield_occ_op}, which leads to:
\begin{eqnarray}
:\! \opcdag{1k}\opc{1k}\! : &=& \cos^2(\theta_{k}) :\! \opfdag{1k}\opf{1k}\! : + \sin^2(\theta_{k}) :\! \opfdag{2k}\opf{2k}\! : \nonumber \\
&+& \cos(\theta_{k}) \sin(\theta_{k}) ( :\! \opfdag{1k}\opf{2k}\! : + \Hc ) \;. \nonumber
\end{eqnarray}
Next, observe that since $|\psi_0\rangle=| \emptyset_1, {\mathrm F}_2\rangle$ we have that
$ :\! \opfdag{1k}\opf{1k}\! : =  \opfdag{1k}\opf{1k}$, 
and $ :\! \opfdag{1k}\opf{2k}\! : =  \opfdag{1k}\opf{2k}$. However, 
\[
 :\! \opfdag{2k}\opf{2k}\! : =  \opfdag{2k}\opf{2k} - 1 = - \opf{2k}\opfdag{2k} \;.  
\]
Hence, in the continuum limit we have:
\begin{eqnarray}
\Delta \rho(x,t) &=& \cos^2(\Theta) \langle \psi(t) | \opfdag{1}(x) \opf{1}(x) | \psi(t) \rangle \\
&-& \sin^2(\Theta) \langle \psi(t) | \opf{2}(x) \opfdag{2}(x) | \psi(t) \rangle \nonumber \\
&+& \cos(\Theta) \sin(\Theta) ( \langle \psi(t) | \opfdag{1}(x) \opf{2}(x) | \psi(t) \rangle + \mathrm{c.c.})
\nonumber
\end{eqnarray}
where we used the short notation $\Theta\equiv\Theta(x)$ for the continuum version of the thermofield angle (see \cref{eq:thermofield_angle}), defined through the Fermi function:
\begin{align}
\sin^2(\Theta(x))  \equiv \frac{1}{\nep^{\beta Wx} + 1} \;.  
\end{align}
The final transformation involves re-writing the $\opf{\chain}(x)$ in terms of orthogonal chain operators, see Eq.~\eqref{eq:orthPol_transform_inv}.
One can show that the quantities involved in the expectation value are all well defined. For instance:
\[
\opf{2}(x) \opfdag{2}(x)= \sum_{n,m=0}^{\infty}
U_{2,n}(x) U_{2,m}(x) \, \opa{2,n} \opadag{2,m} \;.
\]
Hence, by considering  $\langle \psi(t) | \opa{2,n} \opadag{2,m} | \psi(t)\rangle$, one 
easily realises that these matrix elements vanish exactly for $n,m > \tilde{L}(t)$, where
$\tilde{L}(t)$ is the effective distance reached by the excitations at time $t$. 
This implies that the infinite sums are all effectively cut-off by $\tilde{L}(t)$.

\begin{figure}[t]
\centering
\includegraphics[width=8.5cm]{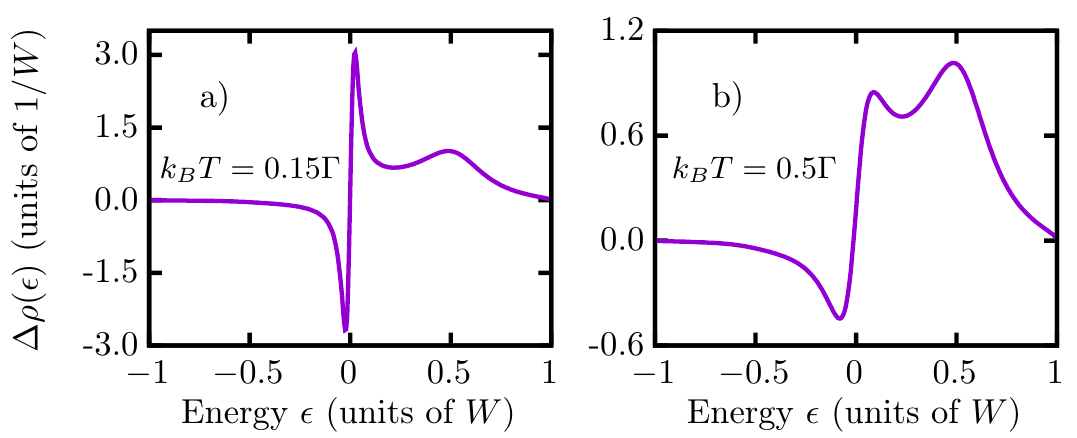}
\caption{Conduction electrons occupation $\Delta\rho(\epsilon,t)$ in the symmetric SIAM with $U=2.5\pi\Gamma$, at temperatures $k_BT=0.15\Gamma$(a) and $k_BT=0.5\Gamma$(b). The initial state $|\psi_0\rangle$ is here a factorized state with doubly occupied impurity 
$|\!\!\uparrow\downarrow\rangle$ and conduction modes in the thermal state, 
$|\psi_0\rangle=|\!\!\uparrow\downarrow\rangle\otimes\fullemptyKet$. 
$\Delta\rho(\epsilon,t)$ is plotted for times $t=15\hbar/\Gamma$ (a) and $t=10\hbar$ (b), after which it does not change anymore.}
\label{fig:BathOccStartFilled}
\end{figure}

\section{Starting from an occupied impurity} \label{sec:startFilled}
We have previously seen that the conduction bath occupation density $\Delta\rho(\epsilon,t)$ shows a peak corresponding to the impurity level $\varepsilon_d$ (see \cref{fig:IntOccup}). 
However, the second impurity level at energy $\varepsilon_d+U$ was not been observed. 
The reason is that such level corresponds to a double occupied state, which plays only a minor role in the dynamics, when starting from an empty impurity. 
Here, we study $\Delta\rho(\epsilon,t)$ for the same dynamics as before, starting, however, from the doubly occupied impurity state $|\!\!\uparrow\downarrow\rangle$, see  \cref{fig:BathOccStartFilled}. We clearly observe the peak close to energy 
$\varepsilon_d+U=1.25\pi\Gamma=0.125\pi W$, while peaks at energy $\varepsilon_d$ --- corresponding to the empty impurity state --- are absent.

\begin{figure}[t]
\centering
\includegraphics[width=8.5cm]{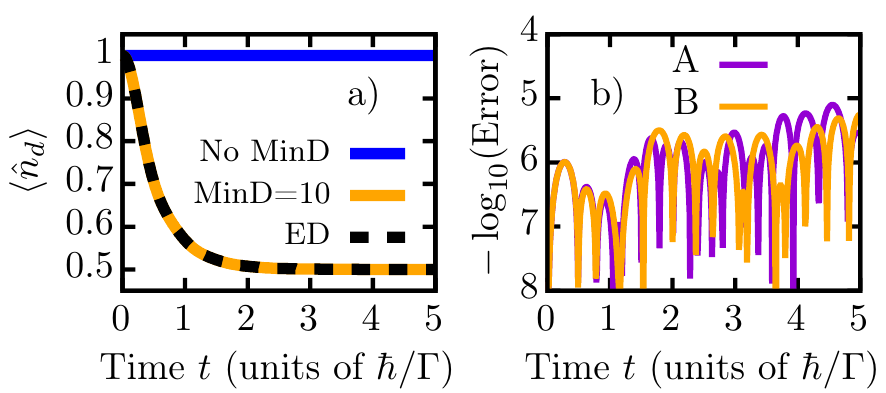}
\caption{a) Impurity occupation obtained when using \textbf{B}, at $U=0$ and temperature $k_BT=0$, for an initially occupied impurity. TDVP gets stuck due to the next-nearest neighbor interaction, as manifested by the constant impurity occupation. The problem is resolved by setting a minimum bond dimension, and the dynamics agrees well with ED data. b) Error of the impurity occupation, calculated as the difference of MPS and ED data for structures \textbf{A} and \textbf{B}, employing a minimum bond dimension MinD=10. Note that in structure \textbf{A} interactions are only nearest neighbor, and thus projection errors are absent.}
\label{fig:TDVP_ImpOcc_Error_MinD}
\end{figure}

\section{TDVP beyond nearest neighbor hopping} \label{sec:minD}
This section is devoted to an analysis of the projection error of the time dependent variational principle to compute the dynamics of the system. Our structure \textbf{A} contains at most nearest neighbor interactions, and thus projection errors are absent when using 2-site TDVP~\cite{AoP_Paeckel_2019}. On the other hand, the Hamiltonians of structures \textbf{B} and \textbf{C} both contain mainly next-nearest neighbor terms, where projection errors do not vanish in general. For simplicity, we restrict our analysis to the noninteracting case $U=0$, where only structures \textbf{A} and \textbf{B} are relevant (structure \textbf{C} is equivalent to \textbf{B}). 
Similarly to the previous section, we initialize the impurity in the filled state 
$|1\rangle$ with one spinless fermion, and set the temperature to $T=0$ to avoid strong entanglement growth in structure \textbf{A}. 

We study the dynamics of the impurity occupation, which for $U=0$ can easily be compared with ED results.
Without setting a minimum bond dimension we find TDVP to get stuck, as indicated by the horizontal curve in \cref{fig:TDVP_ImpOcc_Error_MinD}. 
Since the impurity is initially filled, the interaction term between impurity and the first filled chain site does not change the state. On the other hand, the impurity electron could move to the first empty chain site. In the MPS, however this interaction is a next-nearest neighbor term, and since the initial state is a product state, this process is projected out by TDVP. 
Setting a minimum bond dimension for the state, we can enlarge the projector to avoid this issue. Indeed we find excellent agreement with ED data for the impurity occupation
(see \cref{fig:TDVP_ImpOcc_Error_MinD}(b))), with error similar to structure \textbf{A} 
(see \cref{fig:TDVP_ImpOcc_Error_MinD}(b)), where projection errors do not play a role.

\section{Convergence with Bond dimension} \label{sec:bond_conv}
\begin{figure}[t]
\centering
\includegraphics[width=8.5cm]{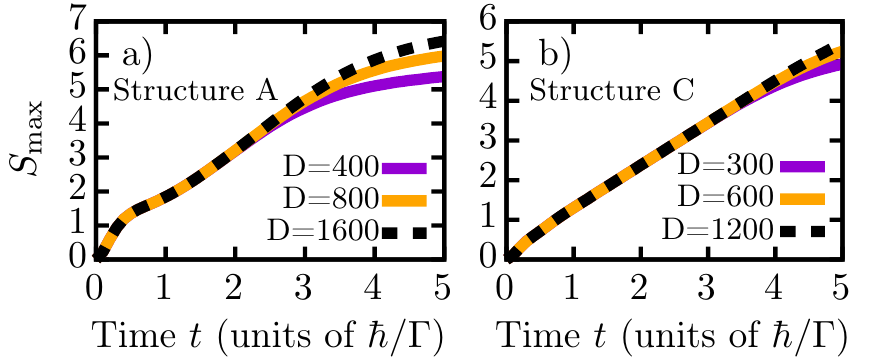}
\caption{Dynamics of the maximum entanglement entropy along the MPS, $S_{\mathrm{max}}$, for different bond dimensions $D$ and Structures \textbf{A} (panel a) and \textbf{C} (panel b), at temperature $k_B=4\Gamma$. Even at bond dimension $D=1600$, the entanglement entropy is clearly not converged at the end of the dynamics for structure \textbf{A}, while structure \textbf{C} is sufficiently converged at $D=1200$.}
\label{fig:EntBondConvergence}
\end{figure}

The bond dimension $D$ is the crucial numerical parameter in our simulations, as it sets an upper bound for the number of states kept. 
To ensure that the simulations deliver correct results we need to converge the quantity of interest with respect to the bond dimension. Here we study the convergence of the maximum entanglement entropy $S_\mathrm{max}$ for the structures \textbf{A} and \textbf{C} 
(see \cref{fig:EntBondConvergence}). 
We omit details for structure \textbf{B} as results where converged already at $D=150$. For structures \textbf{A} and \textbf{C} instead, we find significantly slower convergence, due to the higher entanglement. For structure \textbf{A} -- separating filled and empty chains -- $S_\mathrm{max}$ is clearly not converged at the end of the simulation even with bond dimension $D=1600$. Indeed the entanglement increase seems to be linear in time, but starts to flatten due to the insufficient bond dimension. However, the massive increase of computational costs prevents us from going to higher $D$. For structure \textbf{C}, instead, we are able to reach convergence using a bond dimension of $D=1200$. Note that this is still significantly larger than bond dimension, $D=150$, required for structure \textbf{B}, and longer simulations would be impossible due to the required exponentially increasing bond dimension.

\bibliographystyle{apsrev4-1}
\bibliography{BiblioMPS_OQS}

\begin{thebibliography}{48}%
\makeatletter
\providecommand \@ifxundefined [1]{%
 \@ifx{#1\undefined}
}%
\providecommand \@ifnum [1]{%
 \ifnum #1\expandafter \@firstoftwo
 \else \expandafter \@secondoftwo
 \fi
}%
\providecommand \@ifx [1]{%
 \ifx #1\expandafter \@firstoftwo
 \else \expandafter \@secondoftwo
 \fi
}%
\providecommand \natexlab [1]{#1}%
\providecommand \enquote  [1]{``#1''}%
\providecommand \bibnamefont  [1]{#1}%
\providecommand \bibfnamefont [1]{#1}%
\providecommand \citenamefont [1]{#1}%
\providecommand \href@noop [0]{\@secondoftwo}%
\providecommand \href [0]{\begingroup \@sanitize@url \@href}%
\providecommand \@href[1]{\@@startlink{#1}\@@href}%
\providecommand \@@href[1]{\endgroup#1\@@endlink}%
\providecommand \@sanitize@url [0]{\catcode `\\12\catcode `\$12\catcode
  `\&12\catcode `\#12\catcode `\^12\catcode `\_12\catcode `\%12\relax}%
\providecommand \@@startlink[1]{}%
\providecommand \@@endlink[0]{}%
\providecommand \url  [0]{\begingroup\@sanitize@url \@url }%
\providecommand \@url [1]{\endgroup\@href {#1}{\urlprefix }}%
\providecommand \urlprefix  [0]{URL }%
\providecommand \Eprint [0]{\href }%
\providecommand \doibase [0]{http://dx.doi.org/}%
\providecommand \selectlanguage [0]{\@gobble}%
\providecommand \bibinfo  [0]{\@secondoftwo}%
\providecommand \bibfield  [0]{\@secondoftwo}%
\providecommand \translation [1]{[#1]}%
\providecommand \BibitemOpen [0]{}%
\providecommand \bibitemStop [0]{}%
\providecommand \bibitemNoStop [0]{.\EOS\space}%
\providecommand \EOS [0]{\spacefactor3000\relax}%
\providecommand \BibitemShut  [1]{\csname bibitem#1\endcsname}%
\let\auto@bib@innerbib\@empty
\bibitem [{\citenamefont {Anderson}(1961)}]{Anderson_PR61}%
  \BibitemOpen
  \bibfield  {author} {\bibinfo {author} {\bibfnamefont {P.~W.}\ \bibnamefont
  {Anderson}},\ }\href {\doibase 10.1103/PhysRev.124.41} {\bibfield  {journal}
  {\bibinfo  {journal} {Phys. Rev.}\ }\textbf {\bibinfo {volume} {124}},\
  \bibinfo {pages} {41} (\bibinfo {year} {1961})}\BibitemShut {NoStop}%
\bibitem [{\citenamefont {Kondo}(1964)}]{Kondo_PTP64}%
  \BibitemOpen
  \bibfield  {author} {\bibinfo {author} {\bibfnamefont {J.}~\bibnamefont
  {Kondo}},\ }\href@noop {} {\bibfield  {journal} {\bibinfo  {journal}
  {Progress of theoretical physics}\ }\textbf {\bibinfo {volume} {32}},\
  \bibinfo {pages} {37} (\bibinfo {year} {1964})}\BibitemShut {NoStop}%
\bibitem [{\citenamefont {Hewson}(1997)}]{Hewson_kondo:book}%
  \BibitemOpen
  \bibfield  {author} {\bibinfo {author} {\bibfnamefont {A.~C.}\ \bibnamefont
  {Hewson}},\ }\href@noop {} {\emph {\bibinfo {title} {The Kondo problem to
  heavy fermions}}}\ (\bibinfo  {publisher} {Cambridge University Press},\
  \bibinfo {year} {1997})\BibitemShut {NoStop}%
\bibitem [{\citenamefont {Metzner}\ and\ \citenamefont
  {Vollhardt}(1989)}]{Metzner_PRL89}%
  \BibitemOpen
  \bibfield  {author} {\bibinfo {author} {\bibfnamefont {W.}~\bibnamefont
  {Metzner}}\ and\ \bibinfo {author} {\bibfnamefont {D.}~\bibnamefont
  {Vollhardt}},\ }\href {\doibase 10.1103/PhysRevLett.62.324} {\bibfield
  {journal} {\bibinfo  {journal} {Phys. Rev. Lett.}\ }\textbf {\bibinfo
  {volume} {62}},\ \bibinfo {pages} {324} (\bibinfo {year} {1989})}\BibitemShut
  {NoStop}%
\bibitem [{\citenamefont {Georges}\ \emph {et~al.}(1996)\citenamefont
  {Georges}, \citenamefont {Kotliar}, \citenamefont {Krauth},\ and\
  \citenamefont {Rozenberg}}]{Georges_RMP96}%
  \BibitemOpen
  \bibfield  {author} {\bibinfo {author} {\bibfnamefont {A.}~\bibnamefont
  {Georges}}, \bibinfo {author} {\bibfnamefont {G.}~\bibnamefont {Kotliar}},
  \bibinfo {author} {\bibfnamefont {W.}~\bibnamefont {Krauth}}, \ and\ \bibinfo
  {author} {\bibfnamefont {M.~J.}\ \bibnamefont {Rozenberg}},\ }\href {\doibase
  10.1103/RevModPhys.68.13} {\bibfield  {journal} {\bibinfo  {journal} {Rev.
  Mod. Phys.}\ }\textbf {\bibinfo {volume} {68}},\ \bibinfo {pages} {13}
  (\bibinfo {year} {1996})}\BibitemShut {NoStop}%
\bibitem [{\citenamefont {Rams}\ and\ \citenamefont
  {Zwolak}(2020)}]{PRL_Rams_2020}%
  \BibitemOpen
  \bibfield  {author} {\bibinfo {author} {\bibfnamefont {M.~M.}\ \bibnamefont
  {Rams}}\ and\ \bibinfo {author} {\bibfnamefont {M.}~\bibnamefont {Zwolak}},\
  }\href {\doibase 10.1103/PhysRevLett.124.137701} {\bibfield  {journal}
  {\bibinfo  {journal} {Phys. Rev. Lett.}\ }\textbf {\bibinfo {volume} {124}},\
  \bibinfo {pages} {137701} (\bibinfo {year} {2020})}\BibitemShut {NoStop}%
\bibitem [{\citenamefont {Schwarz}\ \emph {et~al.}(2018)\citenamefont
  {Schwarz}, \citenamefont {Weymann}, \citenamefont {von Delft},\ and\
  \citenamefont {Weichselbaum}}]{PRL_Weichselbaum_2018}%
  \BibitemOpen
  \bibfield  {author} {\bibinfo {author} {\bibfnamefont {F.}~\bibnamefont
  {Schwarz}}, \bibinfo {author} {\bibfnamefont {I.}~\bibnamefont {Weymann}},
  \bibinfo {author} {\bibfnamefont {J.}~\bibnamefont {von Delft}}, \ and\
  \bibinfo {author} {\bibfnamefont {A.}~\bibnamefont {Weichselbaum}},\ }\href
  {\doibase 10.1103/PhysRevLett.121.137702} {\bibfield  {journal} {\bibinfo
  {journal} {Phys. Rev. Lett.}\ }\textbf {\bibinfo {volume} {121}},\ \bibinfo
  {pages} {137702} (\bibinfo {year} {2018})}\BibitemShut {NoStop}%
\bibitem [{\citenamefont {Reimann}\ and\ \citenamefont
  {Manninen}(2002)}]{RMP_Reimann_2002}%
  \BibitemOpen
  \bibfield  {author} {\bibinfo {author} {\bibfnamefont {S.~M.}\ \bibnamefont
  {Reimann}}\ and\ \bibinfo {author} {\bibfnamefont {M.}~\bibnamefont
  {Manninen}},\ }\href {\doibase 10.1103/RevModPhys.74.1283} {\bibfield
  {journal} {\bibinfo  {journal} {Rev. Mod. Phys.}\ }\textbf {\bibinfo {volume}
  {74}},\ \bibinfo {pages} {1283} (\bibinfo {year} {2002})}\BibitemShut
  {NoStop}%
\bibitem [{\citenamefont {Braun}\ \emph {et~al.}(2004)\citenamefont {Braun},
  \citenamefont {K\"onig},\ and\ \citenamefont {Martinek}}]{PRB_Braun_2004}%
  \BibitemOpen
  \bibfield  {author} {\bibinfo {author} {\bibfnamefont {M.}~\bibnamefont
  {Braun}}, \bibinfo {author} {\bibfnamefont {J.}~\bibnamefont {K\"onig}}, \
  and\ \bibinfo {author} {\bibfnamefont {J.}~\bibnamefont {Martinek}},\ }\href
  {\doibase 10.1103/PhysRevB.70.195345} {\bibfield  {journal} {\bibinfo
  {journal} {Phys. Rev. B}\ }\textbf {\bibinfo {volume} {70}},\ \bibinfo
  {pages} {195345} (\bibinfo {year} {2004})}\BibitemShut {NoStop}%
\bibitem [{\citenamefont {Gull}\ \emph {et~al.}(2011)\citenamefont {Gull},
  \citenamefont {Millis}, \citenamefont {Lichtenstein}, \citenamefont
  {Rubtsov}, \citenamefont {Troyer},\ and\ \citenamefont
  {Werner}}]{RMP_Gull_2011}%
  \BibitemOpen
  \bibfield  {author} {\bibinfo {author} {\bibfnamefont {E.}~\bibnamefont
  {Gull}}, \bibinfo {author} {\bibfnamefont {A.~J.}\ \bibnamefont {Millis}},
  \bibinfo {author} {\bibfnamefont {A.~I.}\ \bibnamefont {Lichtenstein}},
  \bibinfo {author} {\bibfnamefont {A.~N.}\ \bibnamefont {Rubtsov}}, \bibinfo
  {author} {\bibfnamefont {M.}~\bibnamefont {Troyer}}, \ and\ \bibinfo {author}
  {\bibfnamefont {P.}~\bibnamefont {Werner}},\ }\href {\doibase
  10.1103/RevModPhys.83.349} {\bibfield  {journal} {\bibinfo  {journal} {Rev.
  Mod. Phys.}\ }\textbf {\bibinfo {volume} {83}},\ \bibinfo {pages} {349}
  (\bibinfo {year} {2011})}\BibitemShut {NoStop}%
\bibitem [{\citenamefont {Rubtsov}\ \emph {et~al.}(2005)\citenamefont
  {Rubtsov}, \citenamefont {Savkin},\ and\ \citenamefont
  {Lichtenstein}}]{PRB_Rubtsov_2005}%
  \BibitemOpen
  \bibfield  {author} {\bibinfo {author} {\bibfnamefont {A.~N.}\ \bibnamefont
  {Rubtsov}}, \bibinfo {author} {\bibfnamefont {V.~V.}\ \bibnamefont {Savkin}},
  \ and\ \bibinfo {author} {\bibfnamefont {A.~I.}\ \bibnamefont
  {Lichtenstein}},\ }\href {\doibase 10.1103/PhysRevB.72.035122} {\bibfield
  {journal} {\bibinfo  {journal} {Phys. Rev. B}\ }\textbf {\bibinfo {volume}
  {72}},\ \bibinfo {pages} {035122} (\bibinfo {year} {2005})}\BibitemShut
  {NoStop}%
\bibitem [{\citenamefont {Werner}\ \emph {et~al.}(2006)\citenamefont {Werner},
  \citenamefont {Comanac}, \citenamefont {de' Medici}, \citenamefont {Troyer},\
  and\ \citenamefont {Millis}}]{PRL_Werner_2006}%
  \BibitemOpen
  \bibfield  {author} {\bibinfo {author} {\bibfnamefont {P.}~\bibnamefont
  {Werner}}, \bibinfo {author} {\bibfnamefont {A.}~\bibnamefont {Comanac}},
  \bibinfo {author} {\bibfnamefont {L.}~\bibnamefont {de' Medici}}, \bibinfo
  {author} {\bibfnamefont {M.}~\bibnamefont {Troyer}}, \ and\ \bibinfo {author}
  {\bibfnamefont {A.~J.}\ \bibnamefont {Millis}},\ }\href {\doibase
  10.1103/PhysRevLett.97.076405} {\bibfield  {journal} {\bibinfo  {journal}
  {Phys. Rev. Lett.}\ }\textbf {\bibinfo {volume} {97}},\ \bibinfo {pages}
  {076405} (\bibinfo {year} {2006})}\BibitemShut {NoStop}%
\bibitem [{\citenamefont {Wilson}(1975)}]{Wilson_RMP75}%
  \BibitemOpen
  \bibfield  {author} {\bibinfo {author} {\bibfnamefont {K.~G.}\ \bibnamefont
  {Wilson}},\ }\href {\doibase 10.1103/RevModPhys.47.773} {\bibfield  {journal}
  {\bibinfo  {journal} {Rev. Mod. Phys.}\ }\textbf {\bibinfo {volume} {47}},\
  \bibinfo {pages} {773} (\bibinfo {year} {1975})}\BibitemShut {NoStop}%
\bibitem [{\citenamefont {Bulla}\ \emph {et~al.}(2008)\citenamefont {Bulla},
  \citenamefont {Costi},\ and\ \citenamefont {Pruschke}}]{RMP_Bulla_2008}%
  \BibitemOpen
  \bibfield  {author} {\bibinfo {author} {\bibfnamefont {R.}~\bibnamefont
  {Bulla}}, \bibinfo {author} {\bibfnamefont {T.~A.}\ \bibnamefont {Costi}}, \
  and\ \bibinfo {author} {\bibfnamefont {T.}~\bibnamefont {Pruschke}},\ }\href
  {\doibase 10.1103/RevModPhys.80.395} {\bibfield  {journal} {\bibinfo
  {journal} {Rev. Mod. Phys.}\ }\textbf {\bibinfo {volume} {80}},\ \bibinfo
  {pages} {395} (\bibinfo {year} {2008})}\BibitemShut {NoStop}%
\bibitem [{\citenamefont {Stadler}\ \emph {et~al.}(2015)\citenamefont
  {Stadler}, \citenamefont {Yin}, \citenamefont {von Delft}, \citenamefont
  {Kotliar},\ and\ \citenamefont {Weichselbaum}}]{PRL_Stadler_2015}%
  \BibitemOpen
  \bibfield  {author} {\bibinfo {author} {\bibfnamefont {K.~M.}\ \bibnamefont
  {Stadler}}, \bibinfo {author} {\bibfnamefont {Z.~P.}\ \bibnamefont {Yin}},
  \bibinfo {author} {\bibfnamefont {J.}~\bibnamefont {von Delft}}, \bibinfo
  {author} {\bibfnamefont {G.}~\bibnamefont {Kotliar}}, \ and\ \bibinfo
  {author} {\bibfnamefont {A.}~\bibnamefont {Weichselbaum}},\ }\href {\doibase
  10.1103/PhysRevLett.115.136401} {\bibfield  {journal} {\bibinfo  {journal}
  {Phys. Rev. Lett.}\ }\textbf {\bibinfo {volume} {115}},\ \bibinfo {pages}
  {136401} (\bibinfo {year} {2015})}\BibitemShut {NoStop}%
\bibitem [{\citenamefont {Bulla}(1999)}]{PRL_Bulla_1999}%
  \BibitemOpen
  \bibfield  {author} {\bibinfo {author} {\bibfnamefont {R.}~\bibnamefont
  {Bulla}},\ }\href {\doibase 10.1103/PhysRevLett.83.136} {\bibfield  {journal}
  {\bibinfo  {journal} {Phys. Rev. Lett.}\ }\textbf {\bibinfo {volume} {83}},\
  \bibinfo {pages} {136} (\bibinfo {year} {1999})}\BibitemShut {NoStop}%
\bibitem [{\citenamefont {\ifmmode~\check{Z}\else \v{Z}\fi{}itko}\ and\
  \citenamefont {Pruschke}(2009)}]{PRB_Zitko_2009}%
  \BibitemOpen
  \bibfield  {author} {\bibinfo {author} {\bibfnamefont {R.}~\bibnamefont
  {\ifmmode~\check{Z}\else \v{Z}\fi{}itko}}\ and\ \bibinfo {author}
  {\bibfnamefont {T.}~\bibnamefont {Pruschke}},\ }\href {\doibase
  10.1103/PhysRevB.79.085106} {\bibfield  {journal} {\bibinfo  {journal} {Phys.
  Rev. B}\ }\textbf {\bibinfo {volume} {79}},\ \bibinfo {pages} {085106}
  (\bibinfo {year} {2009})}\BibitemShut {NoStop}%
\bibitem [{\citenamefont {Deng}\ \emph {et~al.}(2013)\citenamefont {Deng},
  \citenamefont {Mravlje}, \citenamefont {\ifmmode~\check{Z}\else
  \v{Z}\fi{}itko}, \citenamefont {Ferrero}, \citenamefont {Kotliar},\ and\
  \citenamefont {Georges}}]{PRL_Zitko_2013}%
  \BibitemOpen
  \bibfield  {author} {\bibinfo {author} {\bibfnamefont {X.}~\bibnamefont
  {Deng}}, \bibinfo {author} {\bibfnamefont {J.}~\bibnamefont {Mravlje}},
  \bibinfo {author} {\bibfnamefont {R.}~\bibnamefont {\ifmmode~\check{Z}\else
  \v{Z}\fi{}itko}}, \bibinfo {author} {\bibfnamefont {M.}~\bibnamefont
  {Ferrero}}, \bibinfo {author} {\bibfnamefont {G.}~\bibnamefont {Kotliar}}, \
  and\ \bibinfo {author} {\bibfnamefont {A.}~\bibnamefont {Georges}},\ }\href
  {\doibase 10.1103/PhysRevLett.110.086401} {\bibfield  {journal} {\bibinfo
  {journal} {Phys. Rev. Lett.}\ }\textbf {\bibinfo {volume} {110}},\ \bibinfo
  {pages} {086401} (\bibinfo {year} {2013})}\BibitemShut {NoStop}%
\bibitem [{\citenamefont {Garc\'{\i}a}\ \emph {et~al.}(2004)\citenamefont
  {Garc\'{\i}a}, \citenamefont {Hallberg},\ and\ \citenamefont
  {Rozenberg}}]{PRL_Hallberg_2004}%
  \BibitemOpen
  \bibfield  {author} {\bibinfo {author} {\bibfnamefont {D.~J.}\ \bibnamefont
  {Garc\'{\i}a}}, \bibinfo {author} {\bibfnamefont {K.}~\bibnamefont
  {Hallberg}}, \ and\ \bibinfo {author} {\bibfnamefont {M.~J.}\ \bibnamefont
  {Rozenberg}},\ }\href {\doibase 10.1103/PhysRevLett.93.246403} {\bibfield
  {journal} {\bibinfo  {journal} {Phys. Rev. Lett.}\ }\textbf {\bibinfo
  {volume} {93}},\ \bibinfo {pages} {246403} (\bibinfo {year}
  {2004})}\BibitemShut {NoStop}%
\bibitem [{\citenamefont {Wolf}\ \emph
  {et~al.}(2014{\natexlab{a}})\citenamefont {Wolf}, \citenamefont {McCulloch},\
  and\ \citenamefont {Schollw\"ock}}]{Wolf_PRB14}%
  \BibitemOpen
  \bibfield  {author} {\bibinfo {author} {\bibfnamefont {F.~A.}\ \bibnamefont
  {Wolf}}, \bibinfo {author} {\bibfnamefont {I.~P.}\ \bibnamefont {McCulloch}},
  \ and\ \bibinfo {author} {\bibfnamefont {U.}~\bibnamefont {Schollw\"ock}},\
  }\href {\doibase 10.1103/PhysRevB.90.235131} {\bibfield  {journal} {\bibinfo
  {journal} {Phys. Rev. B}\ }\textbf {\bibinfo {volume} {90}},\ \bibinfo
  {pages} {235131} (\bibinfo {year} {2014}{\natexlab{a}})}\BibitemShut
  {NoStop}%
\bibitem [{\citenamefont {Wolf}\ \emph
  {et~al.}(2014{\natexlab{b}})\citenamefont {Wolf}, \citenamefont {McCulloch},
  \citenamefont {Parcollet},\ and\ \citenamefont
  {Schollw\"ock}}]{PRB_Wolf_2014_Cheb}%
  \BibitemOpen
  \bibfield  {author} {\bibinfo {author} {\bibfnamefont {F.~A.}\ \bibnamefont
  {Wolf}}, \bibinfo {author} {\bibfnamefont {I.~P.}\ \bibnamefont {McCulloch}},
  \bibinfo {author} {\bibfnamefont {O.}~\bibnamefont {Parcollet}}, \ and\
  \bibinfo {author} {\bibfnamefont {U.}~\bibnamefont {Schollw\"ock}},\ }\href
  {\doibase 10.1103/PhysRevB.90.115124} {\bibfield  {journal} {\bibinfo
  {journal} {Phys. Rev. B}\ }\textbf {\bibinfo {volume} {90}},\ \bibinfo
  {pages} {115124} (\bibinfo {year} {2014}{\natexlab{b}})}\BibitemShut
  {NoStop}%
\bibitem [{\citenamefont {Wolf}\ \emph {et~al.}(2015)\citenamefont {Wolf},
  \citenamefont {Go}, \citenamefont {McCulloch}, \citenamefont {Millis},\ and\
  \citenamefont {Schollw\"ock}}]{PRX_Wolf_2015}%
  \BibitemOpen
  \bibfield  {author} {\bibinfo {author} {\bibfnamefont {F.~A.}\ \bibnamefont
  {Wolf}}, \bibinfo {author} {\bibfnamefont {A.}~\bibnamefont {Go}}, \bibinfo
  {author} {\bibfnamefont {I.~P.}\ \bibnamefont {McCulloch}}, \bibinfo {author}
  {\bibfnamefont {A.~J.}\ \bibnamefont {Millis}}, \ and\ \bibinfo {author}
  {\bibfnamefont {U.}~\bibnamefont {Schollw\"ock}},\ }\href {\doibase
  10.1103/PhysRevX.5.041032} {\bibfield  {journal} {\bibinfo  {journal} {Phys.
  Rev. X}\ }\textbf {\bibinfo {volume} {5}},\ \bibinfo {pages} {041032}
  (\bibinfo {year} {2015})}\BibitemShut {NoStop}%
\bibitem [{\citenamefont {Ganahl}\ \emph {et~al.}(2015)\citenamefont {Ganahl},
  \citenamefont {Aichhorn}, \citenamefont {Evertz}, \citenamefont
  {Thunstr\"om}, \citenamefont {Held},\ and\ \citenamefont
  {Verstraete}}]{PRB_Verstraete_2015}%
  \BibitemOpen
  \bibfield  {author} {\bibinfo {author} {\bibfnamefont {M.}~\bibnamefont
  {Ganahl}}, \bibinfo {author} {\bibfnamefont {M.}~\bibnamefont {Aichhorn}},
  \bibinfo {author} {\bibfnamefont {H.~G.}\ \bibnamefont {Evertz}}, \bibinfo
  {author} {\bibfnamefont {P.}~\bibnamefont {Thunstr\"om}}, \bibinfo {author}
  {\bibfnamefont {K.}~\bibnamefont {Held}}, \ and\ \bibinfo {author}
  {\bibfnamefont {F.}~\bibnamefont {Verstraete}},\ }\href {\doibase
  10.1103/PhysRevB.92.155132} {\bibfield  {journal} {\bibinfo  {journal} {Phys.
  Rev. B}\ }\textbf {\bibinfo {volume} {92}},\ \bibinfo {pages} {155132}
  (\bibinfo {year} {2015})}\BibitemShut {NoStop}%
\bibitem [{\citenamefont {Bauernfeind}\ \emph {et~al.}(2017)\citenamefont
  {Bauernfeind}, \citenamefont {Zingl}, \citenamefont {Triebl}, \citenamefont
  {Aichhorn},\ and\ \citenamefont {Evertz}}]{PRX_Bauernfeind_2017}%
  \BibitemOpen
  \bibfield  {author} {\bibinfo {author} {\bibfnamefont {D.}~\bibnamefont
  {Bauernfeind}}, \bibinfo {author} {\bibfnamefont {M.}~\bibnamefont {Zingl}},
  \bibinfo {author} {\bibfnamefont {R.}~\bibnamefont {Triebl}}, \bibinfo
  {author} {\bibfnamefont {M.}~\bibnamefont {Aichhorn}}, \ and\ \bibinfo
  {author} {\bibfnamefont {H.~G.}\ \bibnamefont {Evertz}},\ }\href {\doibase
  10.1103/PhysRevX.7.031013} {\bibfield  {journal} {\bibinfo  {journal} {Phys.
  Rev. X}\ }\textbf {\bibinfo {volume} {7}},\ \bibinfo {pages} {031013}
  (\bibinfo {year} {2017})}\BibitemShut {NoStop}%
\bibitem [{\citenamefont {Linden}\ \emph {et~al.}(2020)\citenamefont {Linden},
  \citenamefont {Zingl}, \citenamefont {Hubig}, \citenamefont {Parcollet},\
  and\ \citenamefont {Schollw\"ock}}]{PRB_Linden_2020}%
  \BibitemOpen
  \bibfield  {author} {\bibinfo {author} {\bibfnamefont {N.-O.}\ \bibnamefont
  {Linden}}, \bibinfo {author} {\bibfnamefont {M.}~\bibnamefont {Zingl}},
  \bibinfo {author} {\bibfnamefont {C.}~\bibnamefont {Hubig}}, \bibinfo
  {author} {\bibfnamefont {O.}~\bibnamefont {Parcollet}}, \ and\ \bibinfo
  {author} {\bibfnamefont {U.}~\bibnamefont {Schollw\"ock}},\ }\href {\doibase
  10.1103/PhysRevB.101.041101} {\bibfield  {journal} {\bibinfo  {journal}
  {Phys. Rev. B}\ }\textbf {\bibinfo {volume} {101}},\ \bibinfo {pages}
  {041101} (\bibinfo {year} {2020})}\BibitemShut {NoStop}%
\bibitem [{\citenamefont {Schollw{\"o}ck}(2005)}]{Schollwoeck_RMP05}%
  \BibitemOpen
  \bibfield  {author} {\bibinfo {author} {\bibfnamefont {U.}~\bibnamefont
  {Schollw{\"o}ck}},\ }\href {\doibase 10.1103/RevModPhys.77.259} {\bibfield
  {journal} {\bibinfo  {journal} {Rev. Mod. Phys.}\ }\textbf {\bibinfo {volume}
  {77}},\ \bibinfo {pages} {259} (\bibinfo {year} {2005})}\BibitemShut
  {NoStop}%
\bibitem [{\citenamefont {Schollw{\"o}ck}(2011)}]{AoP_Schollwoeck_2011}%
  \BibitemOpen
  \bibfield  {author} {\bibinfo {author} {\bibfnamefont {U.}~\bibnamefont
  {Schollw{\"o}ck}},\ }\href {\doibase 10.1016/j.aop.2010.09.012} {\bibfield
  {journal} {\bibinfo  {journal} {Annals of Physics}\ }\textbf {\bibinfo
  {volume} {326}},\ \bibinfo {pages} {96} (\bibinfo {year} {2011})}\BibitemShut
  {NoStop}%
\bibitem [{\citenamefont {Eisert}\ \emph {et~al.}(2010)\citenamefont {Eisert},
  \citenamefont {Cramer},\ and\ \citenamefont {Plenio}}]{RMP_Plenio_2010}%
  \BibitemOpen
  \bibfield  {author} {\bibinfo {author} {\bibfnamefont {J.}~\bibnamefont
  {Eisert}}, \bibinfo {author} {\bibfnamefont {M.}~\bibnamefont {Cramer}}, \
  and\ \bibinfo {author} {\bibfnamefont {M.~B.}\ \bibnamefont {Plenio}},\
  }\href {\doibase 10.1103/RevModPhys.82.277} {\bibfield  {journal} {\bibinfo
  {journal} {Rev. Mod. Phys.}\ }\textbf {\bibinfo {volume} {82}},\ \bibinfo
  {pages} {277} (\bibinfo {year} {2010})}\BibitemShut {NoStop}%
\bibitem [{\citenamefont {Kohn}\ and\ \citenamefont
  {Santoro}(2020)}]{ArXiv_Kohn_2020}%
  \BibitemOpen
  \bibfield  {author} {\bibinfo {author} {\bibfnamefont {L.}~\bibnamefont
  {Kohn}}\ and\ \bibinfo {author} {\bibfnamefont {G.~E.}\ \bibnamefont
  {Santoro}},\ }\href@noop {} {\bibfield  {journal} {\bibinfo  {journal}
  {arXiv:2012.01424}\ } (\bibinfo {year} {2020})}\BibitemShut {NoStop}%
\bibitem [{\citenamefont {Verstraete}\ \emph {et~al.}(2004)\citenamefont
  {Verstraete}, \citenamefont {Garc\'{\i}a-Ripoll},\ and\ \citenamefont
  {Cirac}}]{PRL_Cirac_2004}%
  \BibitemOpen
  \bibfield  {author} {\bibinfo {author} {\bibfnamefont {F.}~\bibnamefont
  {Verstraete}}, \bibinfo {author} {\bibfnamefont {J.~J.}\ \bibnamefont
  {Garc\'{\i}a-Ripoll}}, \ and\ \bibinfo {author} {\bibfnamefont {J.~I.}\
  \bibnamefont {Cirac}},\ }\href {\doibase 10.1103/PhysRevLett.93.207204}
  {\bibfield  {journal} {\bibinfo  {journal} {Phys. Rev. Lett.}\ }\textbf
  {\bibinfo {volume} {93}},\ \bibinfo {pages} {207204} (\bibinfo {year}
  {2004})}\BibitemShut {NoStop}%
\bibitem [{\citenamefont {Zwolak}\ and\ \citenamefont
  {Vidal}(2004)}]{PRL_Vidal_2004}%
  \BibitemOpen
  \bibfield  {author} {\bibinfo {author} {\bibfnamefont {M.}~\bibnamefont
  {Zwolak}}\ and\ \bibinfo {author} {\bibfnamefont {G.}~\bibnamefont {Vidal}},\
  }\href {\doibase 10.1103/PhysRevLett.93.207205} {\bibfield  {journal}
  {\bibinfo  {journal} {Phys. Rev. Lett.}\ }\textbf {\bibinfo {volume} {93}},\
  \bibinfo {pages} {207205} (\bibinfo {year} {2004})}\BibitemShut {NoStop}%
\bibitem [{\citenamefont {Takahashi}\ and\ \citenamefont
  {Umezawa}(1975)}]{ColPhen_Umezawa_1975}%
  \BibitemOpen
  \bibfield  {author} {\bibinfo {author} {\bibfnamefont {Y.}~\bibnamefont
  {Takahashi}}\ and\ \bibinfo {author} {\bibfnamefont {H.}~\bibnamefont
  {Umezawa}},\ }\href {\doibase 10.1142/S0217979296000817} {\bibfield
  {journal} {\bibinfo  {journal} {Collective Phenomena}\ }\textbf {\bibinfo
  {volume} {2}},\ \bibinfo {pages} {55} (\bibinfo {year} {1975})}\BibitemShut
  {NoStop}%
\bibitem [{\citenamefont {de~Vega}\ and\ \citenamefont
  {Ba\~nuls}(2015)}]{PRA_Vega_2015}%
  \BibitemOpen
  \bibfield  {author} {\bibinfo {author} {\bibfnamefont {I.}~\bibnamefont
  {de~Vega}}\ and\ \bibinfo {author} {\bibfnamefont {M.-C.}\ \bibnamefont
  {Ba\~nuls}},\ }\href {\doibase 10.1103/PhysRevA.92.052116} {\bibfield
  {journal} {\bibinfo  {journal} {Phys. Rev. A}\ }\textbf {\bibinfo {volume}
  {92}},\ \bibinfo {pages} {052116} (\bibinfo {year} {2015})}\BibitemShut
  {NoStop}%
\bibitem [{\citenamefont {N\"u\ss{}eler}\ \emph {et~al.}(2020)\citenamefont
  {N\"u\ss{}eler}, \citenamefont {Dhand}, \citenamefont {Huelga},\ and\
  \citenamefont {Plenio}}]{PRB_Plenio_2020}%
  \BibitemOpen
  \bibfield  {author} {\bibinfo {author} {\bibfnamefont {A.}~\bibnamefont
  {N\"u\ss{}eler}}, \bibinfo {author} {\bibfnamefont {I.}~\bibnamefont
  {Dhand}}, \bibinfo {author} {\bibfnamefont {S.~F.}\ \bibnamefont {Huelga}}, \
  and\ \bibinfo {author} {\bibfnamefont {M.~B.}\ \bibnamefont {Plenio}},\
  }\href {\doibase 10.1103/PhysRevB.101.155134} {\bibfield  {journal} {\bibinfo
   {journal} {Phys. Rev. B}\ }\textbf {\bibinfo {volume} {101}},\ \bibinfo
  {pages} {155134} (\bibinfo {year} {2020})}\BibitemShut {NoStop}%
\bibitem [{\citenamefont {Gautschi}(1994)}]{ACM_Gautschi_1994}%
  \BibitemOpen
  \bibfield  {author} {\bibinfo {author} {\bibfnamefont {W.}~\bibnamefont
  {Gautschi}},\ }\href@noop {} {\bibfield  {journal} {\bibinfo  {journal} {ACM
  Transactions on Mathematical Software (TOMS)}\ }\textbf {\bibinfo {volume}
  {20}},\ \bibinfo {pages} {21} (\bibinfo {year} {1994})}\BibitemShut {NoStop}%
\bibitem [{\citenamefont {Chin}\ \emph {et~al.}(2010)\citenamefont {Chin},
  \citenamefont {Rivas}, \citenamefont {Huelga},\ and\ \citenamefont
  {Plenio}}]{JMath_Chin_2010}%
  \BibitemOpen
  \bibfield  {author} {\bibinfo {author} {\bibfnamefont {A.~W.}\ \bibnamefont
  {Chin}}, \bibinfo {author} {\bibfnamefont {{\'A}.}~\bibnamefont {Rivas}},
  \bibinfo {author} {\bibfnamefont {S.~F.}\ \bibnamefont {Huelga}}, \ and\
  \bibinfo {author} {\bibfnamefont {M.~B.}\ \bibnamefont {Plenio}},\
  }\href@noop {} {\bibfield  {journal} {\bibinfo  {journal} {Journal of
  Mathematical Physics}\ }\textbf {\bibinfo {volume} {51}},\ \bibinfo {pages}
  {092109} (\bibinfo {year} {2010})}\BibitemShut {NoStop}%
\bibitem [{\citenamefont {Prior}\ \emph {et~al.}(2010)\citenamefont {Prior},
  \citenamefont {Chin}, \citenamefont {Huelga},\ and\ \citenamefont
  {Plenio}}]{PRL_Prior_2010}%
  \BibitemOpen
  \bibfield  {author} {\bibinfo {author} {\bibfnamefont {J.}~\bibnamefont
  {Prior}}, \bibinfo {author} {\bibfnamefont {A.~W.}\ \bibnamefont {Chin}},
  \bibinfo {author} {\bibfnamefont {S.~F.}\ \bibnamefont {Huelga}}, \ and\
  \bibinfo {author} {\bibfnamefont {M.~B.}\ \bibnamefont {Plenio}},\
  }\href@noop {} {\bibfield  {journal} {\bibinfo  {journal} {Physical review
  letters}\ }\textbf {\bibinfo {volume} {105}},\ \bibinfo {pages} {050404}
  (\bibinfo {year} {2010})}\BibitemShut {NoStop}%
\bibitem [{\citenamefont {Bulla}\ \emph {et~al.}(1997)\citenamefont {Bulla},
  \citenamefont {Pruschke},\ and\ \citenamefont {Hewson}}]{JPCM_Bulla_1997}%
  \BibitemOpen
  \bibfield  {author} {\bibinfo {author} {\bibfnamefont {R.}~\bibnamefont
  {Bulla}}, \bibinfo {author} {\bibfnamefont {T.}~\bibnamefont {Pruschke}}, \
  and\ \bibinfo {author} {\bibfnamefont {A.~C.}\ \bibnamefont {Hewson}},\
  }\href {\doibase 10.1088/0953-8984/9/47/014} {\bibfield  {journal} {\bibinfo
  {journal} {Journal of Physics: Condensed Matter}\ }\textbf {\bibinfo {volume}
  {9}},\ \bibinfo {pages} {10463} (\bibinfo {year} {1997})}\BibitemShut
  {NoStop}%
\bibitem [{\citenamefont {Schr\"oder}\ and\ \citenamefont
  {Chin}(2016)}]{PRB_Schroeder_2016}%
  \BibitemOpen
  \bibfield  {author} {\bibinfo {author} {\bibfnamefont {F.~A. Y.~N.}\
  \bibnamefont {Schr\"oder}}\ and\ \bibinfo {author} {\bibfnamefont {A.~W.}\
  \bibnamefont {Chin}},\ }\href {\doibase 10.1103/PhysRevB.93.075105}
  {\bibfield  {journal} {\bibinfo  {journal} {Phys. Rev. B}\ }\textbf {\bibinfo
  {volume} {93}},\ \bibinfo {pages} {075105} (\bibinfo {year}
  {2016})}\BibitemShut {NoStop}%
\bibitem [{\citenamefont {Gautschi}(2004)}]{Gautschi_2004}%
  \BibitemOpen
  \bibfield  {author} {\bibinfo {author} {\bibfnamefont {W.}~\bibnamefont
  {Gautschi}},\ }\href@noop {} {\emph {\bibinfo {title} {Orthogonal
  polynomials}}}\ (\bibinfo  {publisher} {Oxford University Press, New York},\
  \bibinfo {year} {2004})\BibitemShut {NoStop}%
\bibitem [{\citenamefont {Haegeman}\ \emph {et~al.}(2011)\citenamefont
  {Haegeman}, \citenamefont {Cirac}, \citenamefont {Osborne}, \citenamefont
  {Pi\ifmmode~\check{z}\else \v{z}\fi{}orn}, \citenamefont {Verschelde},\ and\
  \citenamefont {Verstraete}}]{PRL_Haegeman_2011}%
  \BibitemOpen
  \bibfield  {author} {\bibinfo {author} {\bibfnamefont {J.}~\bibnamefont
  {Haegeman}}, \bibinfo {author} {\bibfnamefont {J.~I.}\ \bibnamefont {Cirac}},
  \bibinfo {author} {\bibfnamefont {T.~J.}\ \bibnamefont {Osborne}}, \bibinfo
  {author} {\bibfnamefont {I.}~\bibnamefont {Pi\ifmmode~\check{z}\else
  \v{z}\fi{}orn}}, \bibinfo {author} {\bibfnamefont {H.}~\bibnamefont
  {Verschelde}}, \ and\ \bibinfo {author} {\bibfnamefont {F.}~\bibnamefont
  {Verstraete}},\ }\href {\doibase 10.1103/PhysRevLett.107.070601} {\bibfield
  {journal} {\bibinfo  {journal} {Phys. Rev. Lett.}\ }\textbf {\bibinfo
  {volume} {107}},\ \bibinfo {pages} {070601} (\bibinfo {year}
  {2011})}\BibitemShut {NoStop}%
\bibitem [{\citenamefont {Haegeman}\ \emph {et~al.}(2016)\citenamefont
  {Haegeman}, \citenamefont {Lubich}, \citenamefont {Oseledets}, \citenamefont
  {Vandereycken},\ and\ \citenamefont {Verstraete}}]{PRB_Haegeman_2016}%
  \BibitemOpen
  \bibfield  {author} {\bibinfo {author} {\bibfnamefont {J.}~\bibnamefont
  {Haegeman}}, \bibinfo {author} {\bibfnamefont {C.}~\bibnamefont {Lubich}},
  \bibinfo {author} {\bibfnamefont {I.}~\bibnamefont {Oseledets}}, \bibinfo
  {author} {\bibfnamefont {B.}~\bibnamefont {Vandereycken}}, \ and\ \bibinfo
  {author} {\bibfnamefont {F.}~\bibnamefont {Verstraete}},\ }\href {\doibase
  10.1103/PhysRevB.94.165116} {\bibfield  {journal} {\bibinfo  {journal} {Phys.
  Rev. B}\ }\textbf {\bibinfo {volume} {94}},\ \bibinfo {pages} {165116}
  (\bibinfo {year} {2016})}\BibitemShut {NoStop}%
\bibitem [{\citenamefont {Lubich}\ \emph {et~al.}(2015)\citenamefont {Lubich},
  \citenamefont {Oseledets},\ and\ \citenamefont
  {Vandereycken}}]{SIAM_Lubich_2015}%
  \BibitemOpen
  \bibfield  {author} {\bibinfo {author} {\bibfnamefont {C.}~\bibnamefont
  {Lubich}}, \bibinfo {author} {\bibfnamefont {I.~V.}\ \bibnamefont
  {Oseledets}}, \ and\ \bibinfo {author} {\bibfnamefont {B.}~\bibnamefont
  {Vandereycken}},\ }\href@noop {} {\bibfield  {journal} {\bibinfo  {journal}
  {SIAM Journal on Numerical Analysis}\ }\textbf {\bibinfo {volume} {53}},\
  \bibinfo {pages} {917} (\bibinfo {year} {2015})}\BibitemShut {NoStop}%
\bibitem [{\citenamefont {Paeckel}\ \emph {et~al.}(2019)\citenamefont
  {Paeckel}, \citenamefont {Köhler}, \citenamefont {Swoboda}, \citenamefont
  {Manmana}, \citenamefont {Schollw\"ock},\ and\ \citenamefont
  {Hubig}}]{AoP_Paeckel_2019}%
  \BibitemOpen
  \bibfield  {author} {\bibinfo {author} {\bibfnamefont {S.}~\bibnamefont
  {Paeckel}}, \bibinfo {author} {\bibfnamefont {T.}~\bibnamefont {Köhler}},
  \bibinfo {author} {\bibfnamefont {A.}~\bibnamefont {Swoboda}}, \bibinfo
  {author} {\bibfnamefont {S.~R.}\ \bibnamefont {Manmana}}, \bibinfo {author}
  {\bibfnamefont {U.}~\bibnamefont {Schollw\"ock}}, \ and\ \bibinfo {author}
  {\bibfnamefont {C.}~\bibnamefont {Hubig}},\ }\href {\doibase
  10.1016/j.aop.2019.167998} {\bibfield  {journal} {\bibinfo  {journal} {Annals
  of Physics}\ }\textbf {\bibinfo {volume} {411}},\ \bibinfo {pages} {167998}
  (\bibinfo {year} {2019})}\BibitemShut {NoStop}%
\bibitem [{\citenamefont {Bauernfeind}\ and\ \citenamefont
  {Aichhorn}(2020)}]{Scipost_Bauernfeind_2020}%
  \BibitemOpen
  \bibfield  {author} {\bibinfo {author} {\bibfnamefont {D.}~\bibnamefont
  {Bauernfeind}}\ and\ \bibinfo {author} {\bibfnamefont {M.}~\bibnamefont
  {Aichhorn}},\ }\href {\doibase 10.21468/SciPostPhys.8.2.024} {\bibfield
  {journal} {\bibinfo  {journal} {SciPost Phys.}\ }\textbf {\bibinfo {volume}
  {8}},\ \bibinfo {pages} {24} (\bibinfo {year} {2020})}\BibitemShut {NoStop}%
\bibitem [{\citenamefont {Kohn}\ \emph {et~al.}(2020)\citenamefont {Kohn},
  \citenamefont {Silvi}, \citenamefont {Gerster}, \citenamefont {Keck},
  \citenamefont {Fazio}, \citenamefont {Santoro},\ and\ \citenamefont
  {Montangero}}]{PRA_Kohn_2020}%
  \BibitemOpen
  \bibfield  {author} {\bibinfo {author} {\bibfnamefont {L.}~\bibnamefont
  {Kohn}}, \bibinfo {author} {\bibfnamefont {P.}~\bibnamefont {Silvi}},
  \bibinfo {author} {\bibfnamefont {M.}~\bibnamefont {Gerster}}, \bibinfo
  {author} {\bibfnamefont {M.}~\bibnamefont {Keck}}, \bibinfo {author}
  {\bibfnamefont {R.}~\bibnamefont {Fazio}}, \bibinfo {author} {\bibfnamefont
  {G.~E.}\ \bibnamefont {Santoro}}, \ and\ \bibinfo {author} {\bibfnamefont
  {S.}~\bibnamefont {Montangero}},\ }\href {\doibase
  10.1103/PhysRevA.101.023617} {\bibfield  {journal} {\bibinfo  {journal}
  {Phys. Rev. A}\ }\textbf {\bibinfo {volume} {101}},\ \bibinfo {pages}
  {023617} (\bibinfo {year} {2020})}\BibitemShut {NoStop}%
\bibitem [{\citenamefont {He}\ and\ \citenamefont
  {Millis}(2017)}]{PRB_Millis_2017}%
  \BibitemOpen
  \bibfield  {author} {\bibinfo {author} {\bibfnamefont {Z.}~\bibnamefont
  {He}}\ and\ \bibinfo {author} {\bibfnamefont {A.~J.}\ \bibnamefont
  {Millis}},\ }\href {\doibase 10.1103/PhysRevB.96.085107} {\bibfield
  {journal} {\bibinfo  {journal} {Phys. Rev. B}\ }\textbf {\bibinfo {volume}
  {96}},\ \bibinfo {pages} {085107} (\bibinfo {year} {2017})}\BibitemShut
  {NoStop}%
\bibitem [{\citenamefont {Fishman}\ \emph {et~al.}(2020)\citenamefont
  {Fishman}, \citenamefont {White},\ and\ \citenamefont
  {Stoudenmire}}]{itensor}%
  \BibitemOpen
  \bibfield  {author} {\bibinfo {author} {\bibfnamefont {M.}~\bibnamefont
  {Fishman}}, \bibinfo {author} {\bibfnamefont {S.~R.}\ \bibnamefont {White}},
  \ and\ \bibinfo {author} {\bibfnamefont {E.~M.}\ \bibnamefont
  {Stoudenmire}},\ }\href@noop {} {\  (\bibinfo {year} {2020})},\ \Eprint
  {http://arxiv.org/abs/arXiv:2007.14822} {arXiv:arXiv:2007.14822} \BibitemShut
  {NoStop}%
\end{thebibliography}%

\end{document}